\documentclass[a4paper,11pt]{article}
\pdfoutput=1 

\usepackage{jheppub} 

\usepackage[T1]{fontenc} 
\usepackage{dcolumn}
\newcolumntype{d}{D{.}{.}{2.3}}
\usepackage{booktabs}
\usepackage{xspace}
\usepackage[export]{adjustbox}
\usepackage{slashed}
\usepackage{subcaption}
\usepackage{xcolor}
\usepackage{relsize}
\newcommand{\as}{\ensuremath{\alpha_s}\xspace}
\newcommand{\gs}{\ensuremath{g_s}\xspace}

\newcommand{\di}{\ensuremath{\mathrm{d}}\xspace}
\newcommand{\epem}{\ensuremath{e^+e^-}\xspace}
\newcommand{\mjj}{\ensuremath{m_{j_1 j_2}}\xspace}
\newcommand{\bs}[1]{\ensuremath{b_{#1}}\xspace}
\newcommand{\Ht}{\ensuremath{H_T}\xspace}
\newcommand{\Hhatt}{\ensuremath{\hat{H}_T}\xspace}
\newcommand{\htt}{\ensuremath{\hat H_T/2}\xspace}
\newcommand{\LL}{LL\xspace}
\newcommand{\LLp}{LL$^+$\xspace}
\newcommand{\NLL}{NLL\xspace}
\newcommand{\NTLO}{NNLO\xspace}
\newcommand{\NLO}{NLO\xspace}
\newcommand{\ca}{\ensuremath{C_{\!A}}\xspace}
\newcommand{\nc}{\ensuremath{N_{\!C}}\xspace}
\newcommand{\nf}{\ensuremath{n_f}\xspace}

\newcommand{\cf}{\ensuremath{C_F}\xspace}
\newcommand{\cg}{\ensuremath{c_\Gamma}\xspace}

\newcommand{\HEJ}{HEJ\xspace}
\newcommand{\HighEJ}{High Energy Jets\xspace}

\newcommand{\muf}{\ensuremath{\mu_{\scriptscriptstyle \mathrm{F}}}\xspace}
\newcommand{\mur}{\ensuremath{\mu_{\scriptscriptstyle \mathrm{R}}}\xspace}

\newcommand{\ds}{\ensuremath{\di\sigma}\xspace}
\newcommand{\s}[2]{\ensuremath{s_{#1#2}}\xspace}
\newcommand{\spa}[2]{\ensuremath{\left<{#1#2}\right>}\xspace}
\newcommand{\spb}[2]{\ensuremath{\left[{#1#2}\right]}\xspace}

\hypersetup{pdftitle=High-energy logarithms and precision at the LHC,pdfdisplaydoctitle}
\title{High-energy logarithms and precision at the LHC}
\author[a]{Jeppe R.~Andersen,}
\author[b]{Sebastian Jaskiewicz,}
\author[c]{Andreas Maier,}
\author[d]{Jennifer M.~Smillie}

\affiliation[a]{Institute for Particle Physics Phenomenology, Department of Physics, University of Durham, South Road, Durham DH1 3LE, UK}
\affiliation[b]{Albert Einstein Center for Fundamental Physics, Institut f\"ur Theoretische Physik, Universit\"at Bern, Sidlerstrasse 5, CH-3012 Bern, Switzerland}
\affiliation[c]{The Henryk Niewodnicza\'{n}ski Institute of Nuclear Physics, ul. Radzikowskiego 152, 31-342 Kraków, Poland}
\affiliation[d]{Higgs Centre for Theoretical Physics, School of Physics and Astronomy,  The University of Edinburgh, Edinburgh EH9 3FD, United Kingdom}

\emailAdd{jeppe.andersen@durham.ac.uk}
\emailAdd{sebastian.jaskiewicz@unibe.ch}
\emailAdd{andreas.maier@ifj.edu.pl}
\emailAdd{j.m.smillie@ed.ac.uk}

\preprint{IPPP/26/70}

\abstract{We address the breakdown of fixed-order perturbation theory due to logarithms in $s/p_t^2$ for processes involving two jets or more at the energies currently explored at the LHC.
  The problem manifests itself in various ways:
    an obvious breakdown of the perturbative series for some kinematic distributions, highly asymmetric scale variations in other distributions, and the need for an artificially large renormalisation scale to even achieve numerical stability.
  As we demonstrate, the problem is solved through the resummation of these high-energy logarithms. The customary choice of a large renormalisation scale of~$\mjj$ offers no systematic resolution.

  This breakdown in fixed order predictions is illustrated for the production of $Z$+dijets.
  We show that resummed predictions obtained using \HighEJ matched to fixed
  order obtain accurate predictions for all the relevant distributions
  investigated in a recent ATLAS
  measurement~\cite{ATLAS:2024vqf}.
}

\begin{document}
\allowdisplaybreaks
\maketitle
\flushbottom

\section{Introduction}
\label{sec:intro}
The results in this paper will demonstrate that high-precision theoretical predictions for standard LHC observables require input not just from fixed higher-order corrections but also a systematic treatment of certain logarithmically enhanced contributions.
We will show that several seemingly unrelated problems of the fixed-order perturbative description are in fact caused by a single source of logarithmic corrections.
The issues can all be resolved by adapting a systematic all-order treatment of these logarithms.

The issues investigated apply to all processes with two (or more) jets in a colour octet exchange. This includes processes with a Born level contribution with two powers of the QCD coupling, so e.g.~the QCD channels of $pp\to JJ$ and $pp\to \{\gamma,Z,W,H\}JJ$ (whether considered on-shell or with decaying bosons).
The universality of the source of perturbative corrections will be discussed in section~\ref{sec:fixedorder}, which will also explain the emergence of the high-energy logarithmic corrections at NLO.
The specific issues that we find include
\begin{enumerate}
\item highly asymmetrical scale variations in kinematic distributions ($\ds/dp_{t,V}$, $\ds/\di\phi_{j_1j_2}$) which themselves are not associated directly with $\mjj$.
\item the choice of a special renormalisation scale for studies of the $\ds/\di\mjj$-distributions at large \mjj, whereas all other observables are studied with a $p_t$-based renormalisation scale~\cite{Ellis:1992en,Currie:2017eqf}.
\end{enumerate}
Both issues turn out to be manifestations of an underlying breakdown of the perturbative stability by logarithmically large coefficients in the perturbative series.
In fact, the perturbative corrections are so large that when evaluated with the standard scale $\mur=\htt$ the cross section can become negative for large \mjj.
This was observed also for $pp\to\gamma JJ$ in~\cite{Andersen:2025zxw}.

We will furthermore demonstrate that the customary scale choice in fixed-order predictions of $\mur=\mjj$ for studies of $\ds/\di\mjj$ at large \mjj profoundly modifies the perturbative predictions:
the UV logarithm has its argument maximised and is promoted to a leading logarithm at large \mjj, meaning the perturbative coefficient of the leading logarithm is modified by an unrelated contribution.
But the scale choice of $\mur=\mjj$ is special because of the existence of the high-energy logarithmic corrections countering the high-energy running of the coupling.
The most worrying consequence of this specific choice of \mur is the fact that any dependence of the perturbative correction on the perturbative coupling itself vanishes in the large \mjj limit.
The dependence of the NLO corrections on any variation of the coupling is greatly reduced as a result of this, leading to further spurious results.

We will argue that a better solution is offered by systematically confronting the issue of all-order large logarithmic corrections.
We show that the systematic resummation of the logarithmic corrections which arise at all orders will resolve the perturbative issues of the \mjj-distribution.
The matching of the high-energy resummation to fixed-order results ensures a description similar to that obtained for \NLO for distributions where the high-energy logarithm is not dominant.
The matching also reduces the scale variation to the level of \NLO, but also inherits the issues left by fixed-order expansion of the high-energy effects.
These remaining issues would be reduced by including NNLO corrections.

Section~\ref{sec:scalechoice} contains a discussion of the issues at large \mjj related to the choice of renormalisation scale $\mur=\mjj$.
Section~\ref{sec:fixedorder} illustrates the emergence and universality of the high-energy logarithm at NLO for all relevant collider processes.
Section~\ref{sec:measurement} contains results calculated at NLO illustrating the concepts discussed in section~\ref{sec:fixedorder} within the context of predictions for $pp\to (Z\to)\epem JJ$ for the analysis presented in~\cite{ATLAS:2024vqf}.
Also presented are results obtained with \HighEJ (\HEJ)~\cite{Andersen:2009nu,Andersen:2011hs,Andersen:2016vkp,Andersen:2019yzo} involving not just an all-order resummation of the logarithmic terms, but also matching to both $pp\to (Z\to)\epem JJ$ at NLO and up to $pp\to (Z\to)\epem 5J$ Born-level scattering.
The setup for the predictions included in this study is available at \href{https://hej.hepforge.org}{https://hej.hepforge.org} where the code for \HEJ can also be obtained.

\section{Choice of renormalisation scale}
\label{sec:scalechoice}
Before discussing the impact of the high-energy logarithm on the perturbative predictions it is necessary to discuss the related issue of the choice of renormalisation scale used for predictions at large \mjj.
This discussion will show how the high-energy logarithm which is so easily exposed in the analytical results for the perturbative coefficients nevertheless has been largely hidden in the discussion of predictions for the LHC.
Of course for the total cross section or observables based on the transverse momenta the high-energy logarithm may not be relevant.\footnote{We will in section~\ref{sec:otherdist} demonstrate that the high-energy logarithm is responsible for very asymmetrical scale variation bands even for observables of $p_t$ and the azimuthal angle between jets.}
For studies of quantities where the contribution at large~\mjj is not important, such as inclusive cross sections, the azimuthal angles between jets or the transverse momenta of jets or bosons, it is customary to choose a scale based on the transverse momenta ($p_{t,1}, \Ht/2$ etc.) on the grounds that these scales indicate the relevant hardness of the interaction and minimises the logarithms from the UV renormalisation.

However, for studies of the distribution in \mjj it has often been argued that a scale based on \mjj should be used instead~\cite{Ellis:1992en,Currie:2017eqf}.
This scale choice makes perfect sense for central jets of large transverse momentum, where $s\sim2|t|\sim2|u|$ and a renormalisation scale of either $p_t$ or \mjj ($\mjj\sim 2p_t$) minimises the logarithms from renormalisation, and all other logarithms in the perturbative coefficient are small.

The argument for this choice of scale in~\cite{Ellis:1992en,Currie:2017eqf} is instead based on minimising the scale dependence of the cross section rather than an argument based on the logarithms arising in the perturbative coefficient:
the scale choice of \mjj is advocated to ensure rapid apparent convergence of the cross section and a reduction in the scale variation at large \mjj.
And one might have expected that choosing $\as(\mjj^2)$ would lead to small corrections for $\mjj\to\infty$ since $\as\to 0$.
However, this is not the case when the perturbative coefficient has terms of $\log(\mjj)$ as will be demonstrated in section~\ref{sec:fixedorder}.
In these cases a renormalisation scale choice of \mjj leads to the perturbative corrections becoming \emph{unperturbative}, in the sense that any dependence on \as vanishes in the large \mjj limit.
While the scale setting argument of~\cite{Ellis:1992en,Currie:2017eqf} obtains positive predictions for large $\mjj/p_t$, it does so at the cost of effectively changing the perturbative behaviour and it does not ensure the perturbatively correct high-energy behaviour.
To understand the problem one needs to understand the high-energy behaviour at NLO.
This will be analysed in the following sections.

\subsection{The standard argument for a scale choice}
\label{sec:stdscalechoice}
The standard argument for scale setting starts by considering the perturbative expansion for the $R$-ratio $R=\sigma(\epem\to\text{hadrons})/\sigma(\epem\to\mu^+\mu^-)$.
The perturbative coefficients depend on a ratio of a single physical scale and the renormalisation scale \mur.
Choosing \mur to minimise the logarithm is well motivated.
From e.g.~ref.~\cite{dEnterria:2022hzv,Huston:2023ofk,ParticleDataGroup:2024cfk}:
\begin{align}
  &\as(\mur^2) = \frac {\alpha_s(Q_0^2)}{1 + b_0 \alpha_s(Q_0^2) \log(\mur^2/Q_0^2) - \frac{b_1}{b_0} \alpha_s(Q_0^2) \log \frac{\alpha_s(\mur^2)(b_0 + b_1 \alpha_s(Q_0^2))}{\alpha_s(Q_0^2)(b_0 + b_1 \alpha_s(\mur^2))}+\cdots}
    \label{eq:runningalpha}
\end{align}
with
\begin{align}
b_0=\frac{11\ca-2\nf}{12\pi},\qquad b_1 = \frac{34\ca^2-8 \cf-40\nf}{48\pi^2},  \label{eq:betacoeffs}\\
\end{align}
The $R$-ratio can then be written as
\begin{align}
&R(Q)=R_{EW}(Q)(1+\delta_\mathrm{QCD}(Q)),\qquad \delta_{\mathrm{QCD}}(Q)=\mathlarger{\mathlarger{\sum}}_{n=1}^\infty \bar{c}_n\!\!\left(\mur^2/Q^2\right) \cdot \left(\frac{\as(\mur^2)}{\pi}\right)^n,\label{eq:Rratio}\\
  &\bar{c}_1(\mur^2/Q^2)=c_1, \quad \bar{c}_2(\mur^2/Q^2)=c_2+\pi \bs0 c_1 \log(\mur^2/Q^2),\ldots\\
  &c_1=1, \quad c_2=1.9857 - 0.1152\nf,\ldots.
\end{align}
The argument for choosing $\mur^2\!=\!Q^2$ is well motivated in avoiding logarithmic corrections in the coefficient functions from the running of the coupling from large ratios between the renormalisation scale and the physical scale $Q$.
Since this is a single-scale process, only the ratio of $Q$ and $\mur$ can appear as an argument in a logarithm.
If summed to all orders\footnote{Usual caveats apply about the non-convergence of the series.} the~\mur dependence should cancel as it does order by order in the expression above.

\boldmath
\subsection{Scale setting for dijet processes at large $\mjj/p_t$}
\unboldmath
\label{sec:scalesettingHEL}
In this section we will discuss two separate issues related to choosing a renormalisation scale $\mur=\mjj$.
Firstly, any stabilisation of the cross section relies on a partial compensation between coefficients of unrelated logarithms.
Secondly, when choosing $\mur=\mjj$ the notion of a perturbative expansion at large \mjj vanishes, since any dependence on the coupling disappears.

\subsubsection{Modification of the perturbative coefficient}
\label{sec:acccancelations}

Processes will in general develop perturbative logarithmic corrections related to the UV renormalisation with coefficients proportional to $b_0$ (at NLO) and arguments of $\mur^2$ over the scales $\s i j$.
As discussed in the introduction to this section, when all scales are similar it is of course entirely reasonable to choose any of the scales as a renormalisation scale. The hadronic dijet production is a multi-scale process and new combinations of arguments can appear as we will demonstrate in section~\ref{sec:universality}.

The problem related to the specific scale setting of $\mur=\mjj$ in studies at large $\mjj$ is of an origin different to the UV renormalisation.
As we will discuss in section ~\ref{sec:universality} (see e.g.~equations~\eqref{eq:finiteBFKL} and \eqref{eq:finitecontratnloveto}) the predicted leading logarithmic coefficients for $\ln(\mjj^2/p_t^2)$  should be independent of running coupling effects.
In fact, the coefficients for the leading logarithm are predicted for any $SU(N)$ gauge theory and remain the same even for conformally invariant $\mathcal{N}=4$~SYM.
The coefficient is independent of $n_f$ or $b_0$.
Anticipating the discussion of section~\ref{sec:fixedorder}, the NLO dijet cross section has the form
\begin{align}
  \hat\sigma^{\text{LO}}_{2j}&\sim C \as^2(\mur^2)\\
    \hat\sigma^{\text{NLO}}_{2j}&\sim \hat\sigma^{\text{LO}}_{2j}\left(1-\as(\mur^2)(C_2 \ca  \log\mjj^2/p_t^2 - 2\bs0 \log(\mur^2/p_t^2)\right),
  \label{eq:NLO}
\end{align}
where we have omitted terms which are subleading at large \mjj but have included the \bs0-terms which may or may not become leading depending on the choice of renormalisation scale.
The coefficients $C$ and $C_2$ depend on the specific phase space cuts.
For a conventional $p_t$-based renormalisation scale choice (where the \bs0 term is fixed in the high-energy limit), a poor perturbative convergence is observed for large $\mjj$, as $C_2 \ca \log\mjj^2/p_t^2 \gg 1$.
Eventually, the cross section prediction even becomes negative within the phase
space relevant for LHC analyses for example that of~\cite{ATLAS:2024vqf}.

Evidently, choosing $\mur = \mjj$ instead changes the coefficient of the high-energy logarithm to $C_2 \ca - 2 \bs0$ and thereby affects the onset of the apparent perturbative breakdown.
However, it should be stressed that the two logarithmic terms have completely different physical origins and any moderation of the logarithmic growth is purely accidental.
Normally, a renormalisation scale is chosen to minimise the logarithms of UV running coupling origin.
When choosing $\mur=\mjj$ instead the mechanism for a reduced sensitivity is a reduction in the contribution at large \mjj as a result of a partial compensation between the high-energy logarithm with one coefficient and the UV logarithm with a different coefficient.
Any reduction in sensitivity here is entirely accidental.
On the one hand, the experimental selection, for example the difference between vetoing and not vetoing additional jets, will only affect $C_2$, but obviously not $\bs0$.
On the other hand, the size of the $\bs0$ term depends on the specific process, i.e.~the jet multiplicity, and the number of active flavours, whereas the coefficient $C_A$ of the genuine high-enery logarithm is process-independent and the same even in conformal theories.
We conclude that the choice $\mur = \mjj$ does not lead to a systematic improvement.
At best, it treats the symptoms by delaying the point from which the NLO cross section will be dominated by the high-energy logarithm.

\boldmath
\subsubsection{Disappearance of $\as(\mjj^2)$ in the high-energy limit}
\label{sec:disas}
\unboldmath

\begin{figure}
  \centering
  \includegraphics[width=0.8\textwidth]{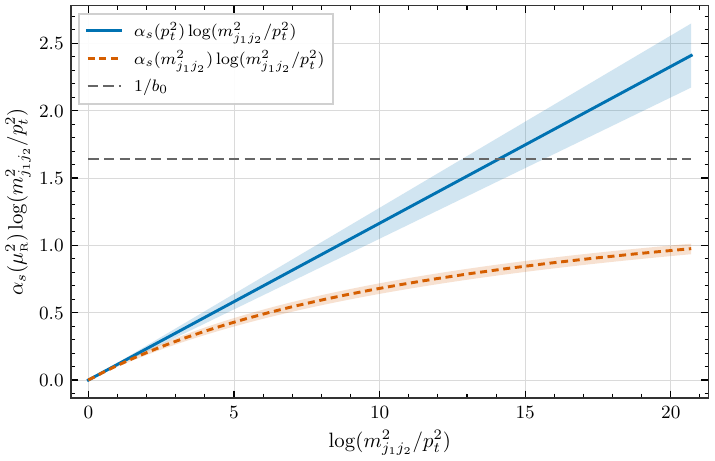}
  \caption{$\as(\mur^2)\log(\mjj^2/p_t^2)$ for a fixed renormalisation scale (blue) and for $\mur=\mjj$ (orange). The choice $\mur=\mjj$ has much reduced \as dependence and a limit which is entirely independent of \as.}
  \label{fig:aslog}
\end{figure}

One might have hoped that choosing $\mur=\mjj$ could help perturbative convergence by forcing $\as(\mur^2)$ to be smaller at large $\mjj$.
However, in this region the product $\as(\mjj^2)\ln(\mjj^2/p_t^2)$ has a special behaviour
\begin{align}
  \as(\mjj^2)\ln(\mjj^2/p_t^2)\to \frac 1 {\bs0}.
  \label{eq:NLOlargemjj}
\end{align}
The behaviour of $\as(\mur^2)\log(\mjj^2/p_t^2)$ for the two choices of \mur is illustrated in figure~\ref{fig:aslog}.
The shaded regions indicate the results under a 10\% variation in the value of $\as(M_Z^2)$.
The impact of this variation is included to illustrate the reduced \as-dependence of the product $\as(\mjj^2)\log(\mjj^2/p_t^2)$ even before the limit of $1/\bs0$ is achieved.

If a renormalisation scale choice is made of $\mur=\mjj$ then the behaviour seen in equation~\eqref{eq:NLO} is fundamentally changed in the high-energy limit:
\begin{align}
  1-\as(\mjj^2)(C_2 \ca - 2\bs0) \log\mjj^2/ p_t^2\to 1-\frac{C_2 \ca-2\bs0}{b_0}.
  \label{eq:highenergykfactor}
\end{align}
The dependence on the coupling has completely disappeared in the high-energy limit, swapped for an explicit $b_0$ dependence.
This form is in contrast to the \emph{conformal} (and $\bs0,\nf$-independent) perturbative behaviour predicted at large $\mjj/p_t$:\footnote{See section~\ref{sec:fixedorder} for a discussion of the perturbative emergence of the high-energy logarithm}
the correction has become \emph{unperturbative} (\as-independent).

\begin{figure}
  \centering
  \includegraphics[width=0.8\textwidth]{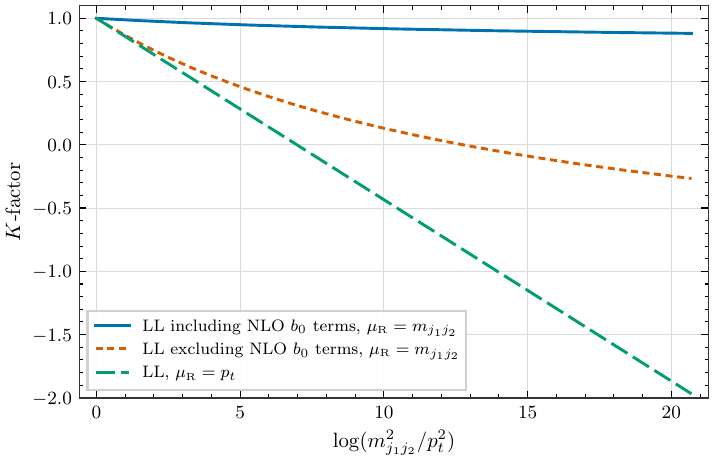}
  \caption{The $K$-factor from equation~(\ref{eq:NLO}) obtained with two choices of \mur. The choice of $\mur=\mjj$ maximises the \bs0-related logarithm and promotes it to leading logarithmic status, thereby changing the coefficient of the high-energy logarithm.}
  \label{fig:kfactor}
\end{figure}
Figure~\ref{fig:kfactor} shows the high-energy approximations of the $K$-factor in equation~\eqref{eq:NLO} achieved for the two choices of $\mur=p_t$ and $\mur=\mjj$. $C_2$ is extracted from the predictions presented in section~\ref{sec:measurement}.
The $K$-factor obtained using $\mur=p_t$ turns negative at around $\mjj=3.2\mathrm{TeV}$.
Choosing $\mur=\mjj$ profoundly modifies the high-energy behaviour through the interplay of running coupling and high-energy terms.

\begin{figure}
  \centering
  \includegraphics[width=0.8\textwidth]{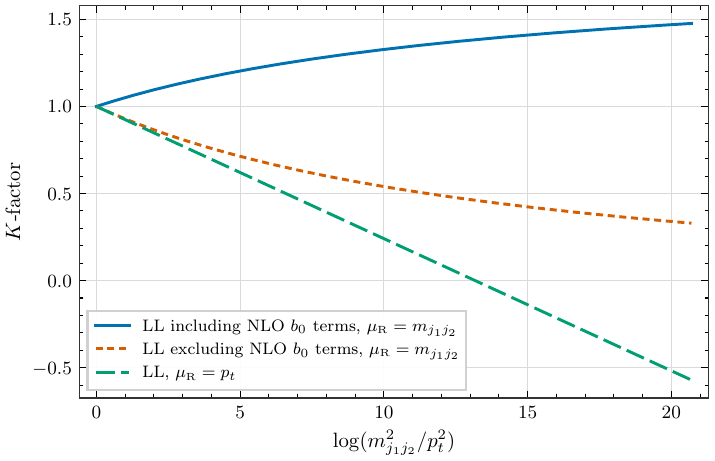}
  \caption{The $K$-factor from equation~(\ref{eq:NLO}) (with parameters corresponding to a situation without a central jet veto) obtained with two choices of \mur. The choice of $\mur=\mjj$ maximises the \bs0-related logarithm and promotes it to leading logarithmic status, thereby changing the coefficient of the high-energy logarithm.}
  \label{fig:kfactor2}
\end{figure}

We recall that the effects of the high-energy logarithm and the running coupling logarithm are unrelated and any compensation is accidental.
This point is illustrated in figure~\ref{fig:kfactor2}, where the $K$-factor has been calculated for the case without a central jet-veto.
In this case the NLO corrections even change sign under the change of the renormalisation scale:
at $\log(\mjj^2/p_t^2)=10$ the corrections change from a 80\% reduction to a 30\% increase.

Evaluating \as at \mjj as suggested in ref.~\cite{Ellis:1992en,Currie:2017eqf} ensures positive cross sections simply through an accidental compensation in the coefficient of logarithms of \mjj introduced by hand by this scale choice, at least for the current setup.

The $K$-factor turning negative does of course not indicate the \emph{start} of the problems for the fixed order description of the perturbative corrections as a function of \mjj, but is the ultimate indication of a problem.
The problem starts as the size of the perturbative corrections increases as a function of \mjj for any of the scale choices.

Since choosing a large $\mur=\mjj$ decreases the value of \as, the impact from the region of large \mjj  on the total cross section and distributions unrelated to \mjj is of course then further suppressed simply because the $\as^2$ Born level contribution is suppressed at large~\mjj.
However, in the case of choosing a renormalisation scale equal to the argument of the leading logarithm the predicted perturbative behaviour is ultimately removed, as discussed around equation~\eqref{eq:highenergykfactor}.

As we will see in section~\ref{sec:CompData}, an alternative to choosing $\mur=\mjj$ to recover positive cross sections at all energies is to systematically deal with the logarithmic corrections to all orders in the coupling.
This approach is based on the factorisation of scattering amplitudes in the high-energy limit as explained in appendix~\ref{sec:corlarges}.
A simple explanation for how this resolves the issue experienced at fixed orders is (as for all resummation methods) that the expression in the bracket of equation~(\ref{eq:NLOlargemjj}) is shown to be universal for all orders and is the first term in the expansion of an exponential.
While the expansion tends towards negative infinity at large \mjj, the exponential is strictly positive for all \mjj.

\section{The high-energy issues with fixed order predictions}
\label{sec:fixedorder}
The recent study~\cite{Andersen:2025zxw} of predictions for the photon+dijets component of the measurement reported in reference~\cite{ATLAS:2024vqf} revealed clearly that the logarithms of high-energy origin are causing issues for fixed-order predictions within the phase space explored by current measurements at the LHC.

We will start this section by showing that the exact same issues are present in $pp\to\mathrm{dijets}+(Z\to)e^+e^-$.
The two processes are of course very similar, and the principle existence of the issues is not surprising.
Nevertheless, it is instructive to explicitly verify the logarithmic dominance and the breakdown of the fixed-order perturbative series within the relevant phase space region.
These effects have previously~\cite{Lindert:2017olm} led to conclusions that the use of central jet vetos cause large instabilities in the studies of $V$+jets.
However, we demonstrate in section~\ref{sec:ppjjZ} that the high-energy logarithmic corrections dominate at fixed-order even without a central jet veto.
Before that, in section~\ref{sec:universality}, we will discuss the universality of the issues at high energies across processes with (at least) two-jets and with a colour octet exchange between these two jets.
It should become clear that high-energy issues  originate from both the universal structure of divergences in one-loop amplitudes and the behaviour of tree-level amplitudes in the limit of high energies, or strictly \emph{Multi-Regge kinematics} (MRK).
Within the field of high-energy corrections the first is called the \emph{Lipatov Ansatz} and the second is embodied in the \emph{Lipatov vertex}.
In section~\ref{sec:universality} we will illustrate their universality from known one-loop results and studies of multi-parton amplitudes.

In studies of dijet-production it has since the very first calculations at NLO~\cite{Ellis:1992en} been customary to choose a large renormalisation scale for studies at large dijet invariant masses, either \mjj directly~\cite{Currie:2017eqf} or a quantity related to it~\cite{Ellis:1992en}.
From the previous discussion it will be clear that the issue necessitating such large scale choice to obtain numerically stable results is in fact of the same high-energy source discussed in section~\ref{sec:universality}.
As discussed in section~\ref{sec:scalechoice}, a large scale choice ensures a numerically stable fixed-order result at high energies, but fails to achieve the correct high-energy behaviour predicted by QCD.

The identified issues are all related to the high-energy logarithms, and the solution is their resummation.
Details are given in appendix~\ref{sec:corlarges}.

\subsection{The emergence and universality of the high-energy logarithm for processes at NLO}
\label{sec:universality}
In this section we will demonstrate explicitly the presence of the high-energy logarithm in the NLO results for both pure dijet production~\cite{Kunszt:1993sd} and then also in the production of $Z +\rm{dijets}$~\cite{Bern:1996ka,Bern:1997sc}.
We will then demonstrate from general one-loop results the universality of the high-energy logarithm at NLO in all processes with a colour-octet exchange between scattered partons using the standard NLO results for the IR poles of amplitudes~\cite{Kunszt:1994np}.
These results are here presented just for one-loop processes using the language of the 4-dimensional on-shell spinor-helicity formalism, but of course they generalise to all orders by the much celebrated Lipatov Ansatz and the wider results of BFKL~\cite{Fadin:1975cb,Kuraev:1976ge,Kuraev:1977fs,Balitsky:1978ic}.
We will start by briefly reviewing the elements necessary for the appearance of the high-energy logarithm in the BFKL formalism and then demonstrate the existence of all such terms explicitly in the NLO results.
For simplicity the discussion here will be limited to leading logarithmic accuracy, but the findings extend to beyond.

\subsubsection{The high-energy logarithm from BFKL}
\label{sec:HEfromBFKL}
In the language of the BFKL literature the Born-level amplitudes for the $2\to2$ QCD processes which dominate in the strict Multi-Regge-Kinematic limit (MRK) are approximated using a factorised form (see e.g.~\cite{DelDuca:2001gu})
\begin{align}
  \mathcal{M}^{(0LL)aa' bb'}_{ij\to ij}=2s \left[\gs\left(T_r^c\right)_{aa'} C^{i(0)}(p_a,p_{a'})\right]\frac 1 t \left[\gs\left(T_r^c\right)_{bb'} C^{j(0)}(p_b,p_{b'})\right],
  \label{eq:bornfact}
\end{align}
where $(0LL)$ signifies the LL component of the Born contribution.
The colour factors $\left(T_r^c\right)_{aa'}$ represent the fundamental or adjoint representation dependent on the parton type in the scattering (quark or gluon, see~\cite{DelDuca:2001gu} for details).
The \emph{impact factors} $C^{i,j}$ can be found in~\cite{Kuraev:1976ge}.
Corrections to this kinematic approximation of the relevant amplitude are suppressed by factors $1/\sqrt{s}$ and so are the amplitudes for the processes which cannot be approximated in this form.
The corrections are therefore suppressed in the MRK limit of $s\gg |t|$, with~$\lvert t\rvert$ fixed.
The point of BFKL is the study of the logarithmically enhanced perturbative corrections to the Born process.
The approximations to the Born amplitude mentioned above are necessary just for obtaining closed analytic results of the impact of the logarithmically enhanced corrections (real and virtual) on the partonic cross section in the MRK limit.
At the cost of giving up closed forms of the inclusive partonic cross sections and integrating the amplitudes numerically, \HEJ uses amplitudes which retain the properties of full QCD/SM amplitudes such as crossing symmetry, gauge invariance, and Lorenz invariance, all of which are missing in amplitudes of the form in eq.~\eqref{eq:bornfact}.

The insight of BFKL is that logarithmic corrections in $s/|t|$ in the cross section arise only in the colour octet channel of the $2\to2$ scattering and with contributions from both the virtual and the real emissions.
We will here describe how this logarithm arises at NLO within BFKL, and then in section~\ref{sec:helogfromnlo} we will demonstrate first how that arises at fixed orders in full QCD and generalises to processes beyond pure QCD.

In standard dimensional regularisation in $D=4-2\varepsilon$ dimensions the contributions from the virtual corrections are described to leading logarithmic accuracy by the simple replacement in equation~\eqref{eq:bornfact} of
\begin{align}
  \frac 1 t &\to \frac 1 t \left(\frac s {-t}\right)^{\alpha(t)}
\end{align}
with
\begin{align}
\alpha(t)&=\gs^2\ \ca\ \frac 2 \varepsilon \left(\frac {\mu^2}{-t}\right)^\varepsilon\ c_\Gamma\\
  c_\Gamma&=\frac 1 {(4\pi)^{2-\varepsilon}}\frac{\Gamma(1+\varepsilon)\ \Gamma^2(1-\varepsilon)}{\Gamma(1-2\varepsilon)}.
\label{eq:alpha}
\end{align}
To next to leading order these are the only virtual corrections relevant for the high-energy leading logarithmic discussion.
Expanded to one-loop it is simply the statement of a logarithmic contribution proportional to $1/\varepsilon$ in the colour octet channel of
\begin{align}
  \mathcal{M}^{(1LL)aa' bb'}_{ij\to ij}=\mathcal{M}^{(0LL)aa' bb'}_{ij\to ij} \alpha(t) \log\left(\frac s {-t}\right).
  \label{eq:oneloopfact}
\end{align}

In addition to the logarithm arising in the virtual corrections, there is a logarithmic correction arising also from the real emission.
This component can be understood from the MRK limit of the square of the $2\to3$ matrix element.
This limit can be derived directly from the analytic form of the spinor helicity amplitudes and the MRK limits of the spinors (see e.g.~\cite{DelDuca:1995zy,DelDuca:1999iql}).
To calculate the contribution we start by noticing that in the MRK limit $s\approx|u|\gg|t|$ and $\log(s/|t|)\approx\log(u/t)$ to leading logarithmic accuracy and $t\approx -p_t^2$ where $p_t$ is the transverse momentum of the final state partons relative to the incoming states.
Finally, we note that the rapidity difference between the final state partons $\Delta y_{ij}=\log(u/t)\approx\log(s/|t|)$ (again to LL accuracy).

The Born level QCD amplitudes are well understood~\cite{Kuraev:1976ge,DelDuca:1995zy} (for all multiplicities) in this limit of all $s_{ij}$ large with fixed transverse momenta, which translates into ordered rapidities $y_1\ll y_2\ll y_3$ but unordered transverse momenta $p_{\perp,1}\sim p_{\perp,2}\sim p_{\perp,3}$.\footnote{The momenta can of course be numbered such that their rapidities are ordered, the LL statement is just that the LL contribution is from the MRK region.}
The limiting behaviour of the colour and spin summed and averaged square of the matrix element is
\begin{align}
  \overline{\left|\mathcal{M}^{(0LL)}_{ij\to igj} \right|^2}=\frac{4 s^2}{\ca^2-1}\frac{\gs^2 C_i}{p_{1\perp}^2}\frac{4\gs^2\ca}{p_{2\perp}^2} \frac{\gs^2 C_j}{p_{3\perp}^2},
\label{eq:sumandavgsquare}
\end{align}
where leading contributions arise only for momentum configurations with the parton content ordered in rapidity (or equivalently light-cone momentum) as indicated.
All other orderings lead to contributions which are power suppressed in $\sqrt{s_{kl}}$ (where $k,l$ signify the relevant pair of partons, see~\cite{Andersen:2017kfc} for further details of the power suppression).
In equation~(\ref{eq:sumandavgsquare}) $C_{l}=\cf$ for quarks and $C_{l}=\ca$ for gluons.

The contribution to the cross section is then
\begin{align}
  \sigma&= \prod_{i=1}^3\int\frac{\di \mathbf{p}_{i\perp}\di y_i}{2 (2\pi)^3} \frac{\overline{\left|\mathcal{M}^{(0LL)}_{ij\to igj} \right|^2}}{s^2} x_a f_{iA}(x_a,Q_a) x_b f_{jB}(x_b,Q_b)\ \delta^2\!\!\left(\sum_{k=1}^3\mathbf{p}_{k\perp} \right)\mathcal{O}_{2j}({p_i}),
          \label{eq:sigmaalpha3}
\end{align}
where the integral is the standard Lorentz-invariant phase space expressed in transverse momenta and rapidity.
Since $\frac{\overline{\left|\mathcal{M}^{(0LL)}_{ij\to igj} \right|^2}}{s^2}$ has no dependence on $y_2$ (to leading logarithmic accuracy in $\left|\mathcal{M}\right|^2,s,x_a$ and $x_b$) for fixed momenta $p_1,p_3$ the integral in equation~(\ref{eq:sigmaalpha3}) is proportional to the range of integration for $y_2$.
The rapidity $y_2$ is integrated between $y_1$ and $y_3$ (since for the leading contribution arises only for this ordering), which according to the observation under equation~(\ref{eq:oneloopfact}) is $\Delta y_{13}\approx \log(s/|t|)$ (to leading logarithmic accuracy).
It is therefore clear that there is a correction proportional to $\alpha_s\log(s/|t|)$ for any choice of $p_1,p_3$ (meaning in particular also for hard emissions $p_{2\perp}\sim p_{l\perp}$, $l=1,3$).

Let us though focus for a moment on the soft region ($p_{2\perp}$ small) to verify the cancellation of the soft pole between real and virtual contributions leaving a finite contribution to the high-energy logarithm.
Integrating $p_{2\perp}$ from 0 to a cut-off $\lambda$ in $D=4-2\varepsilon$ dimensions gives
\begin{align}
  \int \frac{\di \mathbf{p}_{2\perp}\di y_2}{(2\pi)^{2-2\varepsilon} 4\pi}\left(\frac{4\gs^2\ca}{\mathbf{p}^2_{2\perp}}\right) \mu^{2\varepsilon} = \frac {4\gs^2\ca}{(2\pi)^{2-2\varepsilon}4\pi}\Delta y_{13}\frac{\pi^{1-\varepsilon}}{\Gamma(1-\varepsilon)}\frac 1 \varepsilon(\lambda^2/\mu^2)^{-\varepsilon}.
  \label{eq:realeps}
\end{align}
The sum of equation~(\ref{eq:realeps}) and twice the real part of the interference between equation~(\ref{eq:oneloopfact}) and the Born amplitude is indeed free of poles in $\varepsilon$ with a finite contribution for $\varepsilon=0$ of
\begin{align}
  \frac{\as \ca}{\pi}\log(\lambda^2/p^2_{1\perp}) \log(s/|t|),
  \label{eq:finitecontratnlo}
\end{align}
identifying $\Delta y_{13}\sim\log(s/|t|)$.
The logarithmic dependence on $\lambda$ is cancelled by the dependence from the remaining integral over $\mathbf{p}_{2\perp}$ but the result listed here is useful.
If $p_{2\perp}$ is integrated to infinity and all contributions to the light-cone momenta of the incoming partons are ignored (technically allowed to leading logarithmic accuracy) then one arrives at the much celebrated BFKL result of a perturbative correction of
\begin{align}
  4\log(2) \frac{\as \ca}{\pi} \log(s/|t|).
\label{eq:finiteBFKL}
\end{align}
This result in fact holds for any $SU(N)$ gauge theory and even for $\mathcal{N}=4$ SYM~\cite{Kotikov:2000pm}.
Crucially, the effect from the high-energy logarithm is scale independent at leading logarithmic accuracy and originates from the conformal sector of QCD.
As we have seen, this is important for the discussion in section~\ref{sec:scalechoice} since it becomes clear that at large $m_{jj}$ the result should not depend on $b_0$.
It is also worth noting that the presence of \ca is not due to a large \nc approximation, but a result of the logarithm arising in the colour octet channel only.

The result in equation~(\ref{eq:finitecontratnlo}) can directly be used to derive the result for the case of a veto on the further jets - which to leading logarithmic accuracy translates into an upper limit on $p_{2\perp}$ of the jet transverse scale.
In the case of such a $p_{\mathrm{veto}}\ll p_{1\perp}$ one finds directly an expected correction of order (see~\cite{Andersen:2003gs} for further details)
\begin{align}
  \frac{\as \ca}{\pi}\log(p^2_{\mathrm{veto}}/p^2_{1\perp}) \log(s/|t|).
  \label{eq:finitecontratnloveto}
\end{align}
Crucially, the result in equation~\eqref{eq:finiteBFKL} is positive, whereas the result in equation~\eqref{eq:finitecontratnloveto} is negative for the relevant choices of momenta scales.
The result will obviously be dependent on cuts (as investigated also in~\cite{Andersen:2001kta}), and the removal of the positive contribution from a NLO real-emission phase space reduces the $K$-factor.

The treatment in this section of the leading logarithmic component of the perturbative corrections to Born contributions generalises to the colour octet (i.e.~spin 1) component of other processes.
This has allowed the tower of LL corrections to be calculated for Standard Model processes of dijet production also in association with a photon~\cite{Andersen:2025zxw}, $Z/\gamma^*$~\cite{Andersen:2016vkp}, $W$~\cite{Andersen:2020yax} or a Higgs Boson~\cite{Andersen:2017kfc,Andersen:2018tnm} (all on-shell or through the decays).
The formalism used for obtaining the amplitudes used for such predictions is discussed in appendix~\ref{sec:corlarges}.

\subsubsection{The high-energy logarithm from NLO}
\label{sec:helogfromnlo}
In this section we will demonstrate the presence in the one-loop results of the elements from the BFKL inspired leading logarithmic discussion of the previous section.
These components will result in the presence of the high-energy logarithm.
We will demonstrate this for pure dijet production, for dijets in association with a $Z$, and finally for the general processes with a colour octet exchange involving the scattering of two partons.
The necessary behaviour of the real emission contribution was demonstrated in~\cite{Andersen:2009nu} for the relevant Standard Model processes, so we need here be concerned only with the one-loop results.
Of course the result for pure dijets was checked at \NTLO up to \NLL accuracy in~\cite{DelDuca:2001gu} so it will be no surprise that the one-loop result contains the \LL pieces.
We will however demonstrate this using the four dimensional on-shell spinor helicity formalism used in Kunszt-Signer-Tr{\'o}cs{\'a}nyi~\cite{Kunszt:1993sd,Kunszt:1994np} which differs from the formalism used in~\cite{DelDuca:2001gu}.
This discussion therefore serves as an alternative introduction to analysing the high-energy behaviour of one-loop results.
The one-loop results for $ZJJ$ will be analysed from the results of Bern-Dixon-Kosower-Weinzierl~\cite{Bern:1996ka,Bern:1997sc}.
Finally, the appearance of the high-energy logarithm for a general process is discussed using the results of~\cite{Kunszt:1994np}.

\boldmath
\paragraph{The high-energy logarithm for $pp\to jj$ from KST dijet formalism.}
\unboldmath
Using the formalism of~\cite{Kunszt:1993sd} the Born level amplitude for the process $0\to\bar q\bar Q Q q$ is written as (see equation~(3.11) of \cite{Kunszt:1993sd})
\begin{align}
  \mathcal{M}^{(0)}(\bar q, \bar Q;Q, q)&=\gs^2\left(\delta_{i_1 i_3}\delta_{i_2 i_4}-\frac 1 \nc\delta_{i_1 i_4}\delta_{i_2 i_3} \right)a_{4;0}(1,2;3,4),\\
  a_{4;0}(1,2;3,4)&=i\frac{\spa12\spb34}{\s14}.
\label{eq:bornKST}
\end{align}
This scattering process will illustrate the appearance of the high-energy logarithm while also being very simple in that only the colour octet channel contributes at Born level.
In the 't~Hooft-Veltman scheme the one-loop correction to this scattering can be expressed as
\begin{align}
  \mathcal{M}^{(1)}(\bar q,\bar Q;Q,q)=\gs^4 \bigg[&\left(\delta_{i_1 i_3}\delta_{i_2 i_4}-\frac 1 \nc\delta_{i_1 i_4}\delta_{i_2 i_3} \right)a_{4;1}(1,2,3,4)\\
  &+\left(\delta_{i_1 i_3}\delta_{i_2 i_4}\right) a_{4;2}(1,2;3,4)\bigg],\nonumber\\
  a_{4;1}(-,-:+,+)&=\cg\ a_{4;0}(-,-;+,+)\ F_{a;1}^{--}(\varepsilon,s_{12},s_{13},s_{14}),\\
  a_{4;2}(-,-:+,+)&=\cg\ a_{4;0}(-,-;+,+)\ F_{a;2}^{--}(\varepsilon,s_{12},s_{13},s_{14}),
\end{align}
with
\begin{align}
   \label{eq:Fa1--}
  F_{a;1}^{--}(\varepsilon,\s12,\s13,\s14) =\\
  \left(-\frac{\mu^2}{\s14}\right)^\varepsilon&\bigg\{\nc\left[\textcolor{red}{-\frac 2 {\varepsilon^2} - \frac 3 \varepsilon}\textcolor{blue}{+\frac{11}{3\varepsilon}}\textcolor{red}{-\frac 2 \varepsilon \log\frac{\s14}{\s12}}+\frac{13}9 + \pi^2 \right]
  +\nf \left[\textcolor{blue}{-\frac 2 {3\varepsilon}}-\frac{10}9\right] \nonumber\\
 &-\frac 1 \nc\bigg[\textcolor{red}{-\frac 2 {\varepsilon^2}-\frac 3 {\varepsilon}-\frac 2 {\varepsilon}\log\frac{\s 1 2}{\s 1 3}}- 8 +\frac 1 2 \frac {\s14}{\s12}\left(1-\frac{\s13}{\s12}\right)\left(\log^2\frac{\s14}{\s12}+\pi^2\right)\nonumber\\
 &+\frac {\s14}{\s12}\log\frac{\s14}{\s13}\bigg]\bigg\}\textcolor{blue}{-\frac 1 \varepsilon \beta_0},\nonumber
\end{align}
\begin{align}
\label{eq:Fa2--}
 F_{a;2}^{--}(\varepsilon,\s12,\s13,\s14)=\\
 \left(-\frac{\mu^2}{\s14}\right)^\varepsilon\frac{\nc^2-1}{\nc}&\left[\textcolor{red}{-\frac 2 {\varepsilon}\log\frac{\s12}{\s13}}+\frac 1 2 \frac {\s14}{\s12}\left(1-\frac {\s13}{\s12}\right)\left(\log^2 \frac{\s14}{\s13}+\pi^2\right)+\frac{\s14}{\s12}\log\frac{\s14}{\s13}\right].\nonumber
\end{align}
The poles of IR origin have been highlighted in red while the ones from UV origin are highlighted in blue.
The $\mu$ dependence in the IR contributions will cancel with a $\mu$-dependence from the real emission contributions.
All the pieces with kinematic dependence from the curly bracket in equation~\eqref{eq:Fa1--} vanish in the high-energy limit except the singular term proportional to $\log\frac{\s14}{\s12}$ in $F_{a;1}^{--}$.
This will constitute the contribution to the high-energy logarithm from the virtual corrections as discussed in section~\ref{sec:HEfromBFKL}.
The only remaining $\mu$-dependence from equation~\eqref{eq:Fa1--} will be of UV origin.
This is the source of the contribution of the $\bs0$-terms in equation~\eqref{eq:NLO}.
If \mur is chosen as \mjj then this term of running coupling origin can of course contribute in the high-energy limit.

The high-energy logarithm predicted from section~\ref{sec:HEfromBFKL} is immediately identified in eq.~(\ref{eq:Fa1--}) by remembering that for the ``all-outgoing'' momentum notation of~\cite{Kunszt:1993sd} $\s14=t$ while $\s12=s$ where $s,t$ are the invariants used in section~\ref{sec:HEfromBFKL}.
All other pieces with kinematic dependence vanish from equations~(\ref{eq:Fa1--})-(\ref{eq:Fa2--}) in the MRK limit.

This demonstrates that the one-loop results indeed contain the elements necessary to generate the high-energy logarithm predicted by BFKL (once combined with the real-emission component as discussed in section~\ref{sec:HEfromBFKL}).

\boldmath
\paragraph{The high-energy logarithm at NLO for $pp\to JJZ$.}
\unboldmath
We will here demonstrate the appearance of the high-energy logarithm in the process $pp\to JJZ$ using the seminal results of Bern, Dixon, Kosower and Weinzierl~\cite{Bern:1996ka,Bern:1997sc}.
One could start the discussion from the simpler results in~\cite{Bern:1996ka} relevant for the process $qQ\to qQ(Z\to)e^+e^-$ but it is instructive to discuss the slightly more involved amplitudes for $qg\to qg(Z\to)e^+e^-$~\cite{Bern:1997sc}.
The first observation for the Born amplitudes is that just as for the case for $qg\to qg$ the (all outgoing) helicity configuration $q^+g^+g^+\bar q^-$ is suppressed in the high-energy limit compared to $q^+g^+g^-\bar q^-$ (denoted $A_6(1^+_q,2^+,3^-,4_{\bar q}^-)$).
The results for the one-loop colour-stripped primitive amplitudes considered in~\cite{Bern:1996ka,Bern:1997sc} can be written as
\begin{align}
  A_6^{(1)}=\gs^2\cg\left(M^{(0)}V + iF\right),
\end{align}
where $V$ is the divergent contribution and $F$ is finite.
The high-energy logarithm arises from the cut-constructable piece which is given by~\cite{Bern:1997sc}
\begin{align}
  V^{cc}=-\frac 1 {\varepsilon^2}\left(\left(\frac{\mu^2}{-\s12}\right)^\varepsilon+\left(\frac{\mu^2}{-\s23}\right)^\varepsilon+\left(\frac{\mu^2}{-\s34}\right)^\varepsilon\right)-\frac 2 \varepsilon\left(\frac{\mu^2}{-\s56}\right)^\varepsilon-4.
\end{align}
To expose the connection to the prediction from the high-energy limit we move the factor~$\left(\frac{\mu^2}{-\s23}\right)^\varepsilon$ outside of the first bracket and find
\begin{align}
  V^{cc}=-\left(\frac{\mu^2}{-\s23}\right)^\varepsilon\left[\frac 1 {\varepsilon^2}\left(\left(\frac{\s23}{\s12}\right)^\varepsilon+1 + \left(\frac{\s23}{\s34}\right)^\varepsilon\right)\right]-\frac 2 \varepsilon \left(\frac{\mu^2}{-\s56}\right)^\varepsilon-4.
  \label{eq:vcc}
\end{align}
The high-energy limit is defined again with the physical momenta of e.g.~1 and 4 having large light-cone components in one direction and 2 and 3 in the opposite.
The invariant considered fixed (small) is \s23, playing the r\^ole of $t$ from the discussion in section~\ref{sec:HEfromBFKL}.
The high-energy component will then be contained in the $\frac 1 \varepsilon$ contribution of the expansion of equation~(\ref{eq:vcc})
\begin{align}
  V^{cc}\supset\left(\frac{\mu^2}{-\s23}\right)^\varepsilon\left[-\frac 1 \varepsilon \left(\log\left(\s23/\s12\right)+\log\left(\s23/\s34\right)\right)\right].
  \label{eq:vccHElimit}
\end{align}
As long as the momentum of the $Z$ or its decay products do not gain light-cone components faster than the outgoing parton of the set $(1,4)$ or $(2,3)$ then to leading logarithmic accuracy the sum of logarithms in equation~(\ref{eq:vccHElimit}) will constitute the factor of 2 needed in the form of equation~(\ref{eq:alpha}) to get a factorisation of the high-energy logarithmic contribution of the virtual corrections in the form of equation~(\ref{eq:oneloopfact}).
The colour factor \ca arises when the sum over colours is performed.

The relevant behaviour of the real emission contribution was demonstrated in~\cite{Andersen:2009nu} to combine with the virtual correction to give a high-energy correction as discussed in section~\ref{sec:HEfromBFKL}.
This demonstrates the presence of the high-energy logarithm directly in the results of NLO for the process $pp\to JJZ$.

\paragraph{The high-energy logarithm in the general case.}
It is of course no coincidence that the virtual corrections relevant for $pp\to JJZ$ have the necessary soft $1/\varepsilon$ structure for generating the virtual component of the high-energy logarithm.
Kunszt, Signer and Tr{\'o}s{\'a}nyi showed that the QCD soft singular terms for loop helicity amplitudes can be written as~\cite{Kunszt:1994np}
\begin{align}
  \begin{split}
    \mathcal{A}_{\mathrm{soft}}^{1-\mathrm{loop}}&(2\to n-2)_{c_1 c_2\ldots c_n}\\
                                                 &= -\sum_{i<j}\left(\frac{g}{4\pi}\right)^2\cg\frac 1 {\varepsilon^2}\left(-\frac{\mu^2}{\s i j}\right)^\varepsilon\sum_{a,c'_i,c'_j}t^a_{c_i c'_i} t^a_{c_j c'_j}\mathcal{A}^{\mathrm{tree}}(2\to n-2)_{c_1\ldots c'_i\ldots c'_j\ldots c_n},
  \end{split}
  \label{eq:softHElimit}
\end{align}
where $t^a_{c_i c'_i}$ is the $SU(3)$ generator in the relevant colour representation of line $i$, i.e.~$t^a_{ij}$ is $1/2\lambda^a_{ij}$ for an outgoing quark, $-1/2\lambda^{a*}_{ij}$ for an outgoing anti-quark and $if_{aij}$ for an outgoing gluon.
Consider now the Multi-Regge-Kinematic limit of all transverse momenta similar and the $s_{kl}$ large for all final state momenta.
Without loss of generality we can pick the positive light-cone direction along the momentum of the incoming parton 1.
Let $k$ and $l$ be the most forward and backward parton respectively.
In the MRK limit $\s k l$ is then the largest invariant mass between final state momenta.
The dominant Born level configuration is the one with no flavour change between incoming and outgoing parton along the positive light-cone (and also the negative)~\cite{DelDuca:1995zy}.
All other configurations are power suppressed at Born level.
The invariant $\s 1 k$ is not growing in the MRK limit and will play the role of~$t$ from the discussion in section~\ref{sec:HEfromBFKL}.
In fact $\s 1 k\to - p_{\perp k}^2$.
Taking the factor $(-\mu^2/\s 1 k)^\varepsilon$ outside of the sum in equation~(\ref{eq:softHElimit}) one obtains a sum of terms $(\s 1 k/\s i j)^\varepsilon$ the smallest of which is with $\s 1 2 $ and $\s k l$.
These two terms will constitute the leading contribution to the high-energy logarithm.
The colour factors in equation~(\ref{eq:softHElimit}) provide the factor of \ca once contracted with the Born contribution.

This concludes our discussion of the emergence of the high-energy logarithm in NLO calculations.
We have demonstrated the emergence of the soft singular term in the virtual corrections as postulated by the Lipatov ansatz.
We have furthermore summarised how these soft singular virtual terms regulate the divergence from the real emission and the combined result is one of a high-energy logarithm with a coefficient.
The real emission contribution will include emissions of $p_{k\perp}\sim p_{1\perp}$ so the coefficient of the high-energy logarithm will depend on e.g.~phase space cuts.

\boldmath
\section{Predictions for ATLAS 13 TeV measurements of $Z$+dijets }
\label{sec:measurement}
\unboldmath

The production of a $Z$ boson in association with jets at the LHC energy of 13TeV has been studied by ATLAS with an integrated luminosity of 140 fb$^{-1}$~\cite{ATLAS:2024vqf}.
We will demonstrate the impact of the perturbative corrections laid bare in section~\ref{sec:fixedorder} and that the next-to-leading order prediction fail in describing the \mjj-distribution because of the high-energy issues.
The issues caused at fixed order by the large high-energy logarithms will be resolved by the systematic resummation implemented in \HEJ.
Finally we include also predictions from both fixed-order NLO and \HEJ for the remaining distributions measured and included in the analysis.
These will demonstrate that the high-energy logarithms still influence the NLO predictions for $\ds/d\phi_{jj}$ and $\ds/d\Delta p_{t,V}$.
In particular the NLO scale variation becomes very asymmetrical for these distributions because of the remaining dramatic influence from \as-variations from the large corrections at large \mjj.
This feature is inherited by \HEJ because of the matching.
Furthermore, the predictions obtained for $\ds/d\phi_{jj}$ and $\ds/d\Delta p_{t,V}$ from NLO and HEJ for this central scale are very similar and give an equally good description of data.

The cuts used in the ATLAS measurement~\cite{ATLAS:2024vqf} for observations in the channel $pp\to \epem JJ$ are listed in table~\ref{tab:cuts}.
\begin{table}
  \centering
  \begin{tabular}{ll}\toprule
    Lepton rapidity & $|y|\le 1.37$ or $1.52\le|y|\le 2.47$\\
    Leading lepton transverse momentum & $p_T>80$~GeV\\
    Sub-leading lepton transverse momentum & $p_T>7$~GeV\\
    Dilepton mass $m_{ll}$ & $m_{ll}\in (66,116) \mathrm{GeV}$\\
    $p_T^{\mathrm{recoil}}$ & $p_T^{\mathrm{recoil}}>200$GeV\\
    Leading jet transverse momentum & $p_{T,j_1}>80$GeV\\
    Sub-leading jet transverse momentum & $p_{T,j_2}>50$GeV\\
    Jet rapidity & $|y_j|<4.4$\\
    Dijet invariant mass \mjj & $\mjj>200$GeV\\
    Rapidity difference between two hardest jets & $|\Delta y_{j_1j_2}|>1$\\
    Jets with rapidity in-between two hardest jets & None with $p_T>30$GeV\\
    Jet definition & anti-$k_t$, $R=0.4$\\\bottomrule
  \end{tabular}
  \caption{Cuts used in the measurement of ref.\cite{ATLAS:2024vqf} and in the analysis presented here.}
  \label{tab:cuts}
\end{table}
$p_T^{\mathrm{recoil}}$ is defined as the transverse momentum of the hadronic recoil of the charged leptons (in order to obtain a consistent observable across observations of processes with leptons, missing $p_t$ and photons).
We remark that the transverse momenta required for the hardest jets are not particularly small.
A smaller requirement would enhance effects of $\log\mjj^2/p_t^2$ further.
Similarly modest is the requirement of a rapidity separation of the two hardest jets of $|\Delta y_{j_1j_2}|>1$.

Three kinematic distributions reported by ATLAS~\cite{ATLAS:2024vqf} are:
\begin{enumerate}
 \item d$\sigma$/d$p_T^Z$ in 10 bins $200$ GeV $< dp_T^Z < 2600$ GeV
 \item d$\sigma$/d$m_{j_1j_2}$ in 14 bins $200$ GeV $< m_{j_1j_2} < 8200$ GeV
 \item d$\sigma$/d$\Delta\phi_{j_1j_2}$ in 20 bins $-1$  $< \Delta\phi_{j_1j_2}/\pi < 1$
\end{enumerate}
In general, the measurements for this channel are well-described by Standard Model predictions including fixed order except for the dijet invariant mass distribution and curious scale variation results for the fixed order predictions.
We will in the following demonstrate that the high-energy logarithm is the single source of all these issues.

\boldmath
\subsection{$pp\to \mathrm{dijets} + (Z\to)e^+ e^-$ at $\mathcal{O}(\as^3\alpha^2)$}
\label{sec:ppjjZ}
\begin{figure}[tb]
  \centering
  \includegraphics[width=0.8\textwidth]{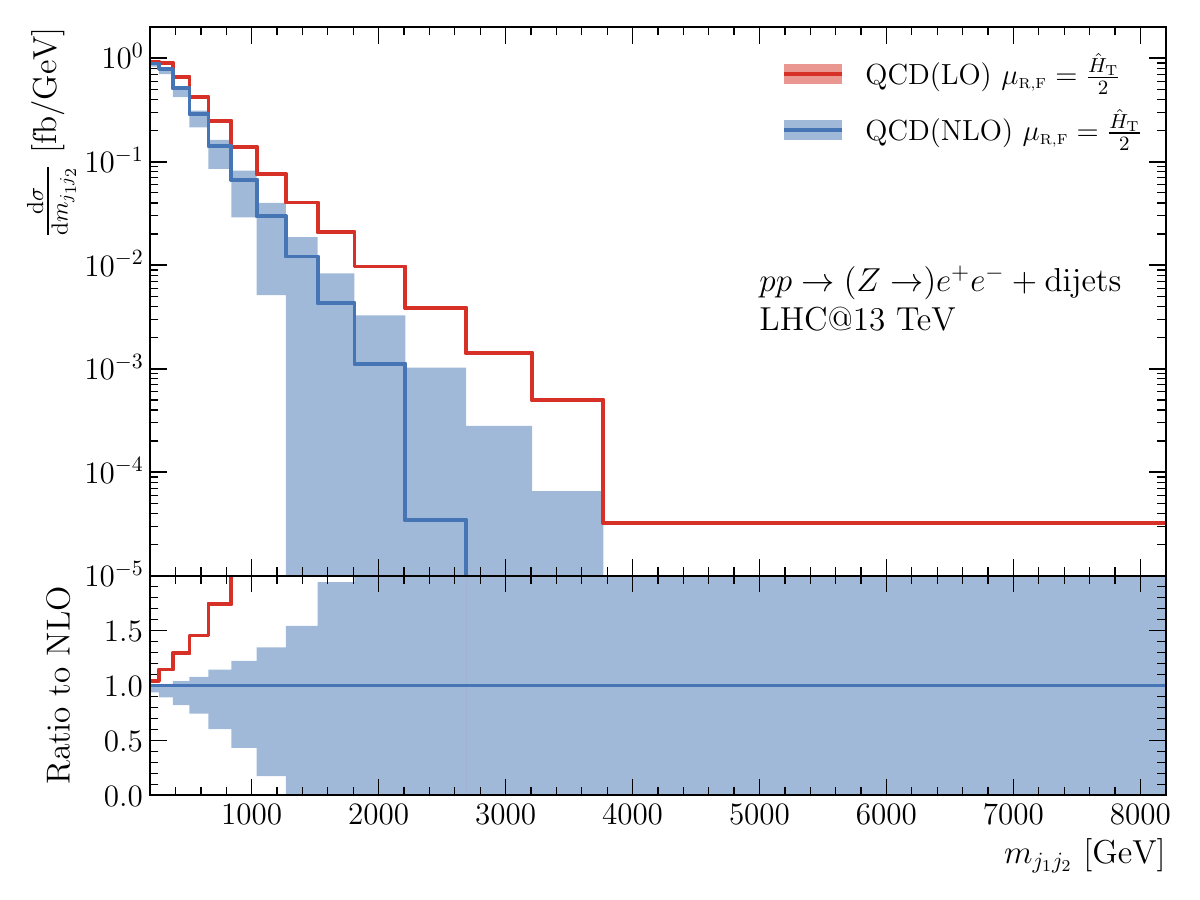}
  \caption{$\di\sigma/\di \mjj$ for LO ($\as^2 \alpha^2$) and NLO (up to $\as^3 \alpha^2$) for $pp\to jj (Z\to)\epem$ with a central scale choice $\mu_R=\mu_F=\htt$ and a standard 7-point scale variation on the result calculated at NLO. }
  \label{fig:fo}
\end{figure}

\unboldmath
Figure~\ref{fig:fo} shows the predictions for $\ds/\di\mjj$ obtained at Born level (using NLO PDFs to expose directly the behaviour of the partonic NLO corrections) and at the next-to-leading order with a central scale choice of $\mur=\htt$ and a standard 7-point scale variation for the NLO result.
The central scale choice of \Hhatt is defined as the scalar sum of the transverse momenta of the jets and $Z$-boson reconstructed from the leptons (i.e.~the vector sum of the two lepton momenta).

The slope of the Born level prediction is largely due to the PDF-determined fall in the parton flux for the increasing $\sqrt{s}$ needed as \mjj increases:
in the large \mjj limit the partonic matrix elements divided by the flux factor depend just on the transverse momenta, but are independent of $\Delta y_{j_1j_2}$ (see reference~\cite{Andersen:2009nu}).

The results at NLO (in blue) represent a large negative correction with the size of the corrections increasing with \mjj.
The results for the scale variation is that a large renormalisation scale (small \as) yields the upper edge (small correction relative to Born), whereas a small renormalisation scale (large \as) results in the lower edge (large correction relative to Born).
One result of the very clear logarithmic dependence of the NLO correction on \mjj is that the scale variation almost vanishes at small \mjj when the logarithmic term is small.
This region has the largest contribution to the total cross section.
But it is clear that the resultant small scale variation in the total cross section is no indication of the scale variation for increasing \mjj (or for the missing higher order terms).

The NLO cross section eventually turns negative for all the choices of renormalisation scales.
For the central scale choice this happens around $\mjj=3$TeV.
This is entirely in line with the expectations from sections~\ref{sec:scalechoice} and~\ref{sec:fixedorder}.
The behaviour is also completely in line with what was observed in~\cite{Andersen:2025zxw} for the fixed-order predictions for the process of dijets production in association with a photon:
the perturbative behaviour is controlled completely by the logarithmic high-energy corrections proportional to $\log(s/p_T^2)$.

The observation is that the logarithmic behaviour is dominant well within the phase space explored at the LHC even for these relatively large transverse momenta of the jets of around $100\mathrm{GeV}$.

\begin{figure}[tb]
  \centering
  \includegraphics[width=0.8\textwidth]{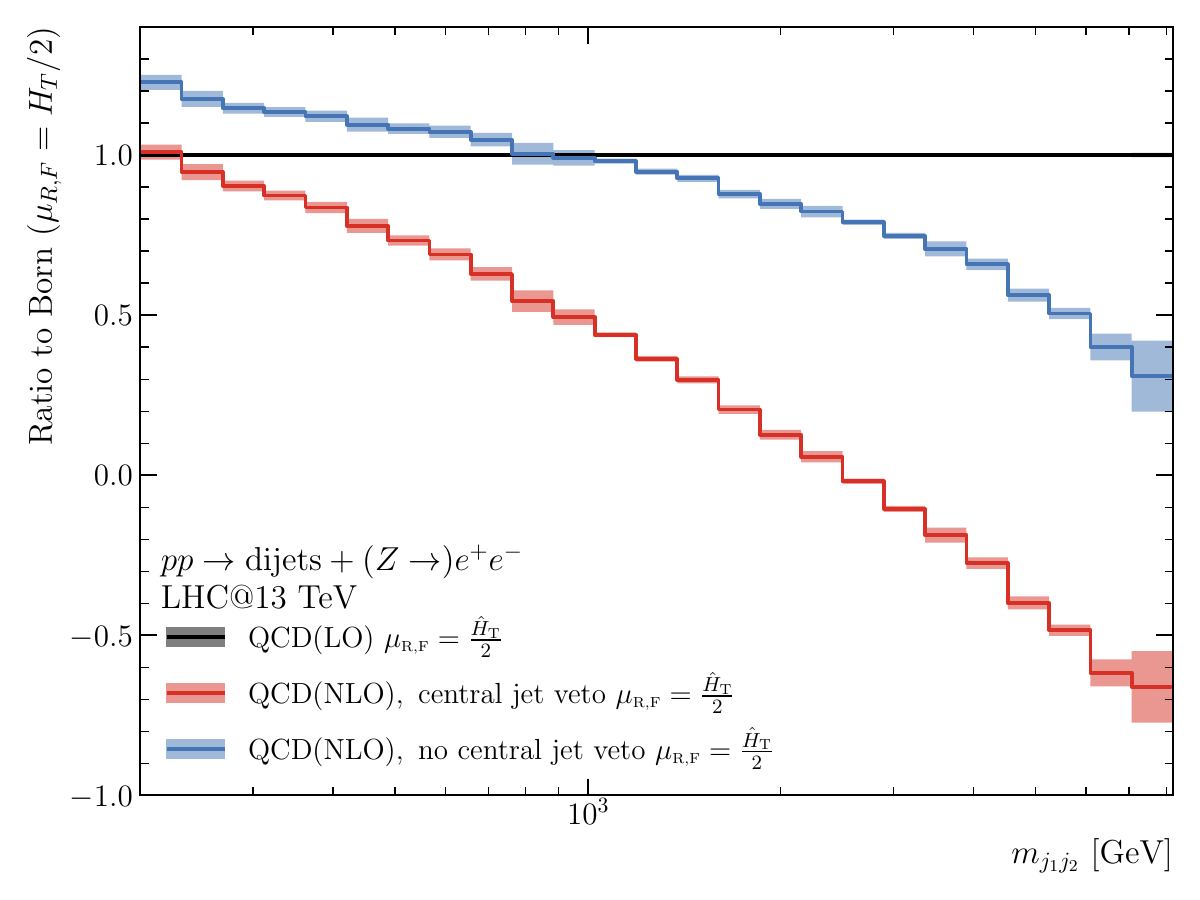}
  \caption{The NLO $K$-factor as a function of $\mjj$ for $pp\to jj (Z\to)\epem$($\as^2 \alpha^2$) with central scale choice $\mu_R=\mu_F=\htt$ with and without a central jet veto. The shaded area indicates the MC associated uncertainty, not the scale variation.}
  \label{fig:NLOKfactor}
\end{figure}

In order to illustrate better the logarithmic dominance of the partonic NLO corrections we plot in figure~\ref{fig:NLOKfactor} in red the ratio between the NLO and LO results (both obtained using the NLO pdfs in order to focus on the partonic NLO corrections).
The width of the curves here indicate the uncertainty from Monte Carlo statistics and not the scale variation.
We use in this plot a linear $y$-axis and a logarithmic $x$-axis and finer binning of equal widths in $\log(\mjj)$.
A pure logarithmic dependence in \mjj of the NLO corrections will display as a straight line.
Such a behaviour is clearly present already from small \mjj for the results (in red) relevant for the predictions for the measurement in~\cite{ATLAS:2024vqf} with the downward trend at large \mjj as the real emission phase space is further constrained by the steeply falling PDFs as $x\to 1$.
This phase space constraint removes a contribution which would otherwise be positive.
The fact that the $K$-factor is close to unity for small \mjj is a coincidence of the cuts used and discussed in the following paragraphs.

The dependence on the perturbative coefficient of the high-energy logarithm was mentioned in section~\ref{sec:HEfromBFKL}.
In particular we mentioned in equation~\eqref{eq:finitecontratnloveto} the logarithmic enhancement of the coefficient when the real-emission contribution is removed with a central jet veto $p_{\perp,\mathrm{veto}}\ll p_{\perp 1}$.
It is therefore natural to ask whether the evident logarithmic behaviour for the analysis is a result of the central jet veto at $p_{\perp,\mathrm{veto}}=30$GeV.

To illustrate the impact of the central jet veto on the logarithmic dependence we include in blue in figure~\ref{fig:NLOKfactor} the result for the $K$-factor without a central jet veto.
The logarithmic dependence clearly also dominates the NLO corrections without a jet veto, but as expected both the sub-leading contributions and the coefficient of the high-energy logarithm are different.
Obviously the NLO $K$-factor is larger without a central jet veto.

While the $K$-factor for the integrated cross section is closer to unity with a jet-veto than without, the absolute value of the coefficient of the high-energy logarithm is largest with a jet veto , which takes the differential $K$-factor further from unity for larger values of \mjj.
This is as expected from the discussion in section~\ref{sec:HEfromBFKL}, although of course the results in figure~\ref{fig:NLOKfactor} are obtained without any kinematic approximations.

The coefficient of the logarithm depends on the specifics of the cuts applied and a specific calculation is necessary to evaluate the impact as is done both in \HEJ and at NLO.
The existence, even the dominance, of the high-energy corrections in the NLO corrections do however not rely on the jet veto.

We note that for a given \mjj the logarithmic influence would obviously be larger if a smaller requirement on the transverse momentum of the hardest jet was used (rather than the 80GeV applied in the current analysis).

This concludes our investigation of the fixed order description of the distribution in~\mjj. We will next discuss the prediction obtained by accounting for the universal high-energy logarithmic contributions to all orders as calculated with \HEJ and compare the results of \HEJ and NLO to data for the distributions measured by ATLAS in~\cite{ATLAS:2024vqf}.

\subsection{All-order predictions using \HEJ and comparisons to data}
\label{sec:CompData}

\begin{figure}[tb]
  \centering
  \includegraphics[width=0.8\textwidth]{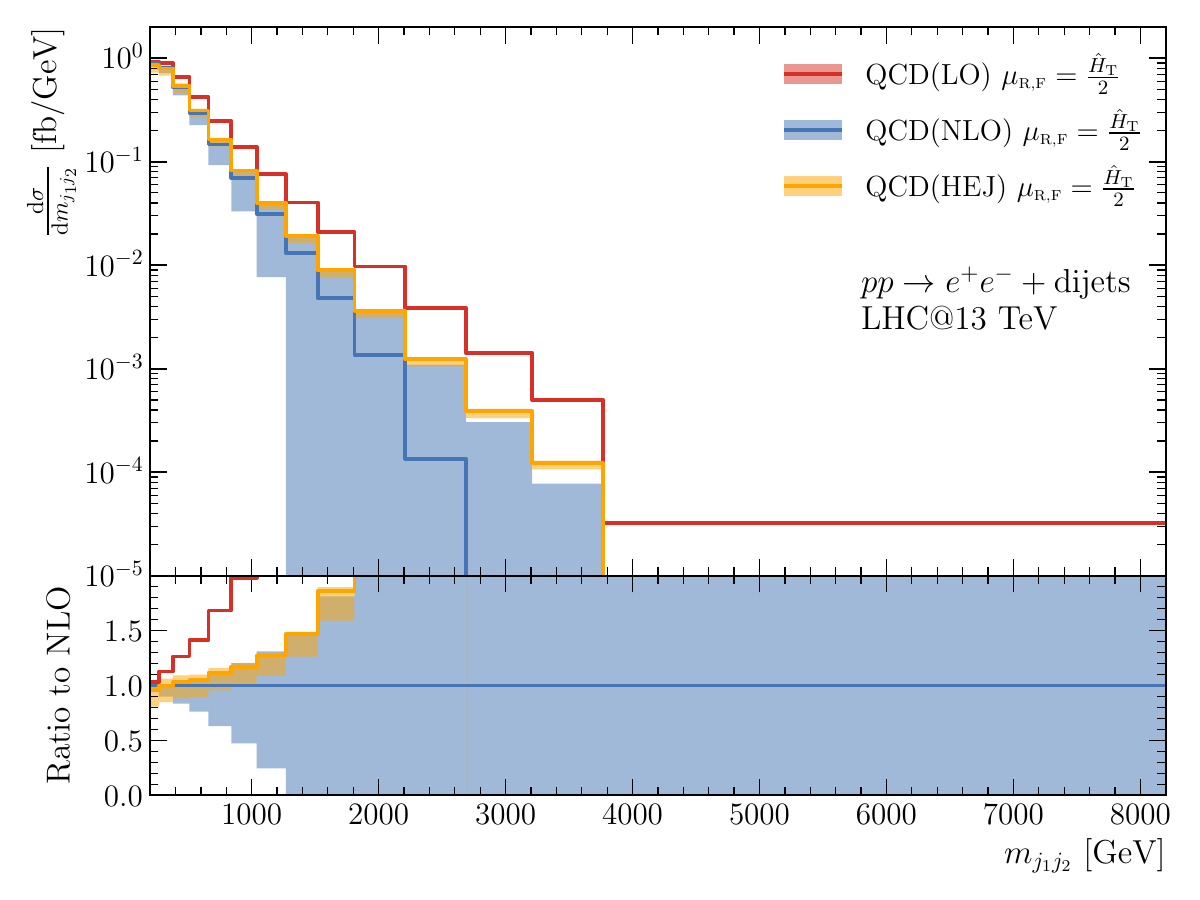}
  \caption{$\di\sigma/\di \mjj$ for \HEJ for $pp\to jj (Z\to)\epem$ (Born: $\as^2 \alpha^2$) with central scale choice $\mu_R=\mu_F=\htt$. For fixed order the variation band is obtained with a standard 7-point scale variation. For HEJ the uncertainty band is obtained using the method described in the text such that the uncertainty band includes the effects of scale variation and of matching.}
  \label{fig:HEJ}
\end{figure}
The result of including the high-energy corrections order by order follows the same procedure described in~\cite{Andersen:2025zxw} (see appendix A of that paper).
The amplitudes used for $Z$+jets in the current study are listed in appendix~\ref{sec:corlarges}.
In addition to the resummation, differential NLO matching is performed multiplicatively bin-by-bin for $\ds/\di p_{t,V}$ and $\ds/\di \Delta \phi_{jj}$, as discussed in ref.~\cite{Andersen:2020yax}.
Such a procedure would obviously not work for $\ds/\di\mjj$ since the distributions expanded to NLO tend to 0 (and beyond).
For this distribution instead we calculate the matching factor for the total cross section between NLO and \HEJ expanded to NLO for each scale choice.
We have checked that the scale variation arrived at this way includes the effect of matching.
We note a curiosity on the scale variation of the NLO result which is a direct result of the dominance of the high-energy logarithmic corrections:
The kinematic matching corrections for the central scale choices are $-10$ to $-30\%$ for $\ds/\di\Delta \phi_{jj}$ and for $\ds/\di p_{t,V}$ they start at less than $-10\%$ decreasing uniformly to $-40\%$ at the largest $p_{t,V}$ of 2.6TeV.

Given the similarities in the processes and cuts it is of course no surprise that the impact of the higher order corrections ends up also being similar for the two processes.
Figure~\ref{fig:HEJ} includes the result for the resummation with \HEJ on top of the fixed-order predictions already included in figure~\ref{fig:fo}.
The all-order result is a large positive correction to NLO and is positive for all \mjj.
This can be a bit overly simplistic and as usual for resummation schemes be understood as a result of NLO retaining only the first term in an expansion of an exponential $\exp(-\as \ca C_2\log\mjj^2/p_t^2)$.
It is of course no surprise either that the difference between keeping just one term or keeping all terms in the expansion of such exponential corrections is smallest when $\as$ is smallest, i.e.~for a large renormalisation scale (upper edge of the NLO scale variation band).
In reality the effects of the cuts and PDFs on the real emission do not exponentiate and the explicit integration over phase space of each multiplicity as in \HEJ is needed.
The shaded region on the results from \HEJ are obtained using a standard 7-point scale variation in all components of the calculation (both resummation and matching) and we checked that the shaded region contains all effects of matching.

Having established that the \mjj-distribution for the QCD component can be meaningfully predicted it becomes interesting to compare to data.
The electroweak component of the cross section should of course be added to the \HEJ results presented in figure~\ref{fig:HEJ} before contrasting the predictions with data.
Since both electroweak and QCD corrections need to be taken into account for the electroweak contribution\footnote{Mixed QCD and electroweak corrections were considered in~\cite{Bargiela:2023npj}, but will not be included in the present study.} we have been using as PDFs the NNPDF3.1luxqed~\cite{NNPDF:2017mvq} which includes the photon PDF in the proton using the luxQED method~\cite{Manohar:2016nzj,Manohar:2017eqh}.
\begin{figure}[tb]
  \centering
  \includegraphics[width=0.8\textwidth]{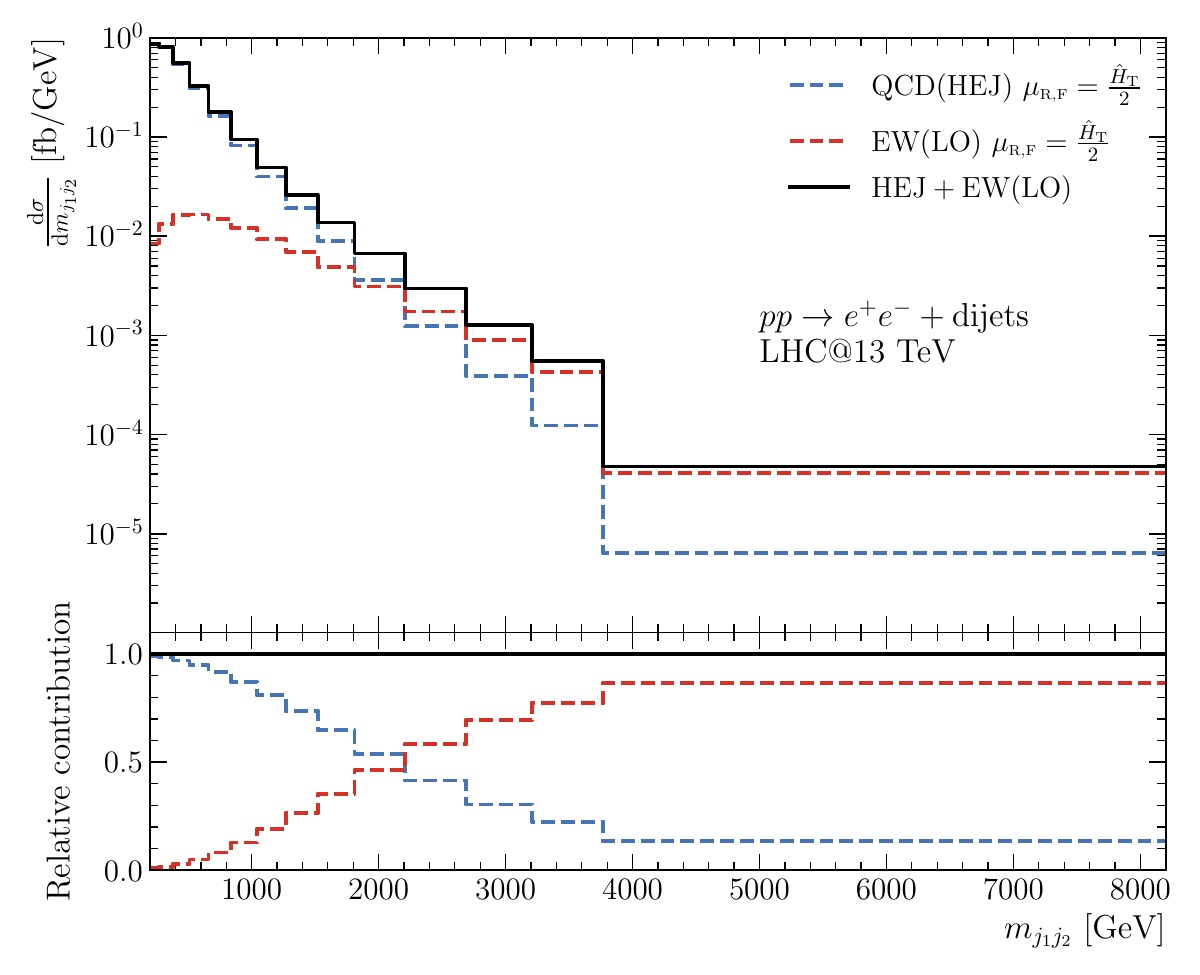}
  \caption{$\di\sigma/\di \mjj$ for \HEJ for $pp\to jj (Z\to)\epem$(Born: $\as^2 \alpha^2$) and LO ($\alpha^4$) with central scale choice $\mu_R=\mu_F=\htt$.}
  \label{fig:HEJvsEW}
\end{figure}
Figure~\ref{fig:HEJvsEW} shows the contribution from the QCD (Born: $\as^2\alpha^2$, resummed and matched with HEJ) process and the electroweak process (Born: $\alpha^4$, evaluated just at Born level) both evaluated with a scale choice of $\mur=\muf=\Hhatt/2$.
The electroweak component dominates the cross section from around 2TeV.
The relative contribution of the electroweak component is significantly larger than in the case of photon production plus dijets.

We calculated the corrections to the EW component for both NLO(QCD) and NLO(EW) (using Sherpa 3.03~\cite{Sherpa:2024mfk}).
In line with the results reported in~\cite{Lindert:2022ejn} we find that the electroweak process receives large NLO corrections for large \mjj from both the electroweak and the QCD sector and that the corrections contribute with opposite sign.
The NLO corrections in the phase space relevant for the measurement in~\cite{ATLAS:2024vqf} are far larger than those relevant for the phase space considered in~\cite{Lindert:2022ejn}.
Since we would like to investigate the impact of the corrections to the electroweak component better in a separate study, we will here restrict the comparison to data of the \mjj-distribution to the region in \mjj where the NLO corrections are not too large and also the electroweak process itself is not dominating.
This is to ensure that it is clearly the description of the QCD component which is being tested against data.
To be precise this means comparisons with data in the region of $\mjj<1.52$~TeV (the exact point chosen to coincide with a bin edge in the measurements).
In this region the QCD process still dominates the cross section and the NLO corrections to the electroweak process are relatively well controlled.
Since the QCD and electroweak NLO correction contribute with opposite signs and we will be comparing only where the contribution suppressed, we will be using just the Born level prediction for the electroweak process.
The sizeable perturbative corrections to the electroweak process are actually likely to reduce the contribution from the electroweak component slightly compared to the Born level prediction.
However, we do not claim an even better description of data until the electroweak corrections have been understood further.

\begin{figure}[tb]
  \centering
  \includegraphics[width=0.48\textwidth]{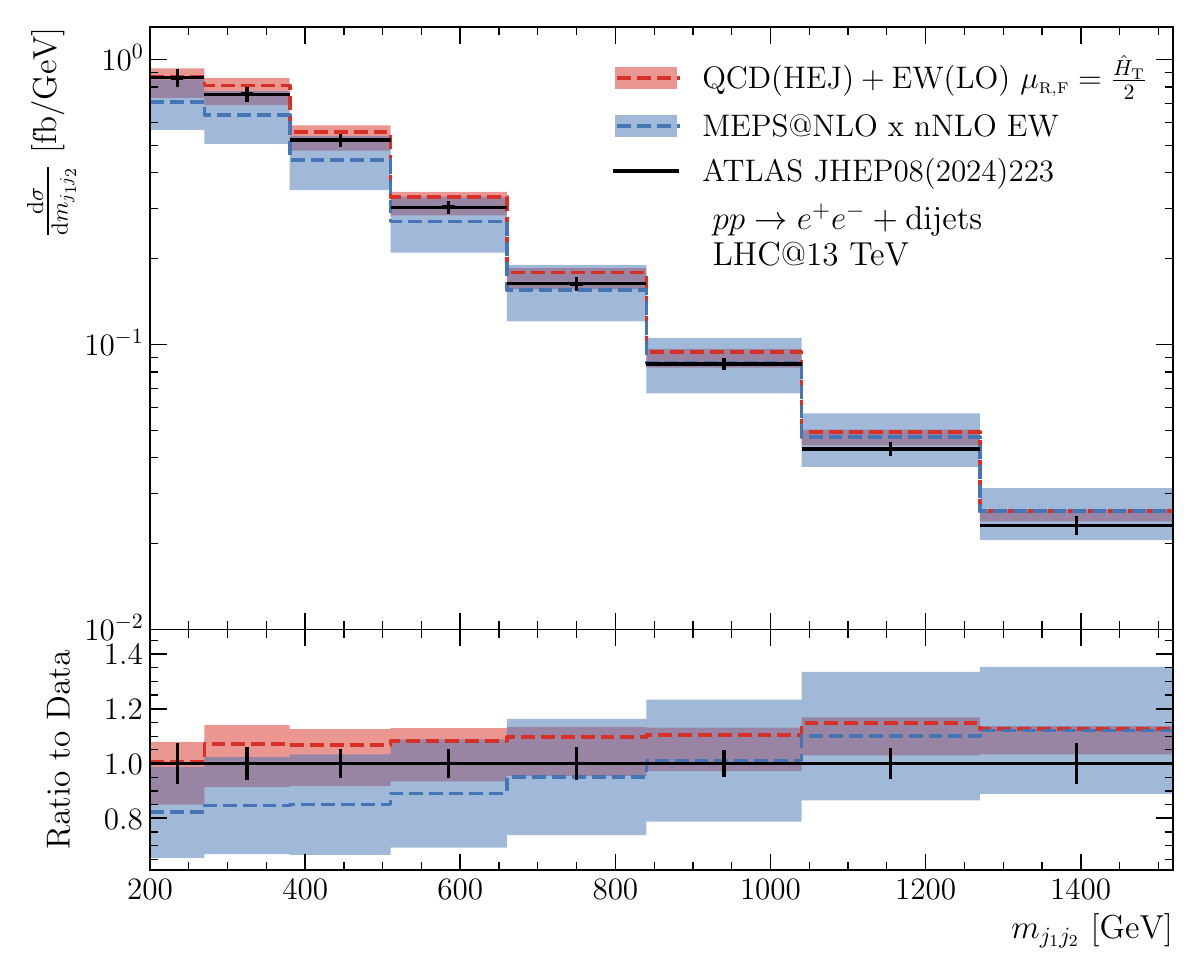}
  \includegraphics[width=0.48\textwidth]{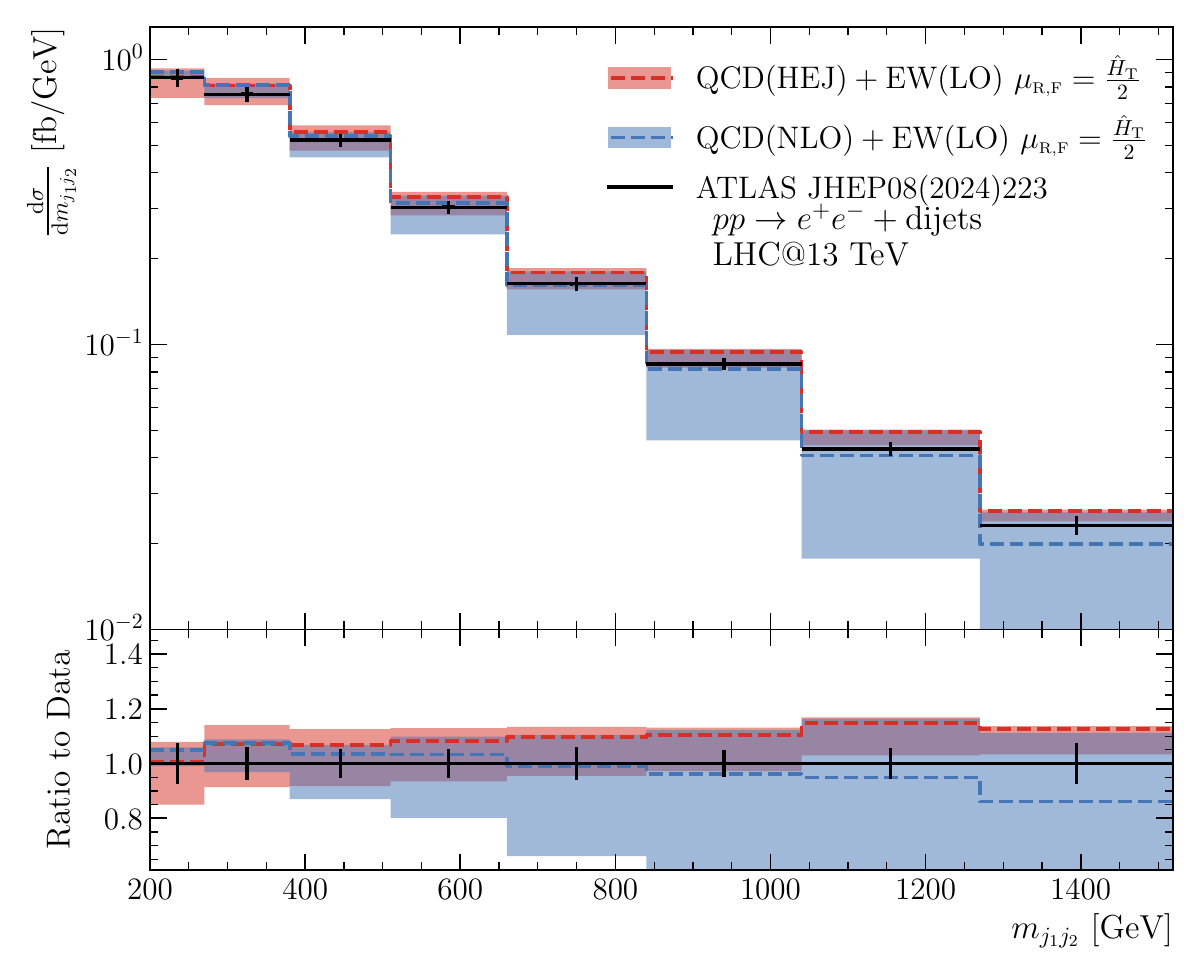}
  \caption{$\di\sigma/\di \mjj$ for \HEJ for $pp\to jj (Z\to)\epem$(Born: $\as^2
    \alpha^2$) with central scale choice $\mu_R=\mu_F=\htt$ and a 7-point scale
    variation. Included also is the prediction from MEPS@NLO taken
    from~\cite{ATLAS:2024vqf} (left) and the QCD(NLO)+EW(LO) prediction (right).}
  \label{fig:HEJvsData}
\end{figure}
Figure~\ref{fig:HEJvsData} shows the prediction from HEJ+EW(LO) compared to data from ATLAS.
Also shown on the left is the prediction presented in~\cite{ATLAS:2024vqf} based
on MEPS@NLO~\cite{Hoeche:2012yf} and, on the right, the QCD(NLO)+EW(LO) prediction. Even if the QCD(NLO) itself is unphysical, the EW(LO) contribution ensures the sum is positive even in the regions where QCD(NLO) is negative.

A good description of data is achieved based on including the high-energy corrections as implemented in \HEJ and the electroweak processes calculated at Born level using Sherpa~\cite{Sherpa:2024mfk}.
These results confirm the conclusion presented in~\cite{Andersen:2025zxw} of the breakdown of the fixed-order perturbative predictions in the study of distributions in \mjj relevant for current LHC energies.
It also confirms the solution in terms of resumming the high-energy logarithmic contributions from both the real and the virtual corrections as first derived by Balitsky, Fadin, Kuraev and Lipatov~\cite{Fadin:1975cb,Kuraev:1976ge,Kuraev:1977fs,Balitsky:1978ic}.

It is perhaps worth reminding the reader that the r\^ole of the resummation through the BFKL equation is a control of the evolution directly in $\log(s)$.
In this sense the slope of the distribution is controlled by the resummation and the high-energy logarithms.
It is therefore not surprising that a control of the high-energy logarithmic corrections is necessary in order to correctly describe the slope of $\di\sigma/\di\mjj$.
Such systematic control is missing in the approach of MEPS@NLO and \NLO descriptions.

Accepting that the all-order treatment and not the fixed-order scale setting of $\mur=\mjj$ is the fix for the issues caused by large logarithmic corrections one might ask how best to apply the fixed-order results.
It is worth remembering here that the resummed cross section obtained with \HEJ is perturbatively matched to the NLO cross section.
This matching procedure can be extended to higher orders by simply terminating the fixed-order calculation of the resummed result at a different order.
The result presented in figure~\ref{fig:HEJvsData} contains the full fixed-order information on the cross section and the resummed and matched description on the distribution.
Increasing the perturbative order of the description of the integrated cross section will lead to smaller scale variation and hopefully a better description of the integrated cross section.

\subsection{Distributions where the high-energy issue at fixed order is masquerading}
\label{sec:otherdist}
We will in this section report on the remaining two distributions investigated by ATLAS~\cite{ATLAS:2024vqf} namely $\ds/\di p_T(\epem)$ and $\ds/\di\Delta\phi_{jj}$.
For such observables the standard practice in fixed-order calculations is to choose a renormalisation scale of \htt.
It turns out that the high-energy logarithm impacts the fixed-order description also of the two remaining distributions measured in~\cite{ATLAS:2024vqf}.
Since the NLO QCD provides physical predictions for these observables it is possible and reasonable to match the HEJ predictions bin by bin in the distributions.

\begin{figure}[tb]
  \centering
  \includegraphics[width=0.48\textwidth]{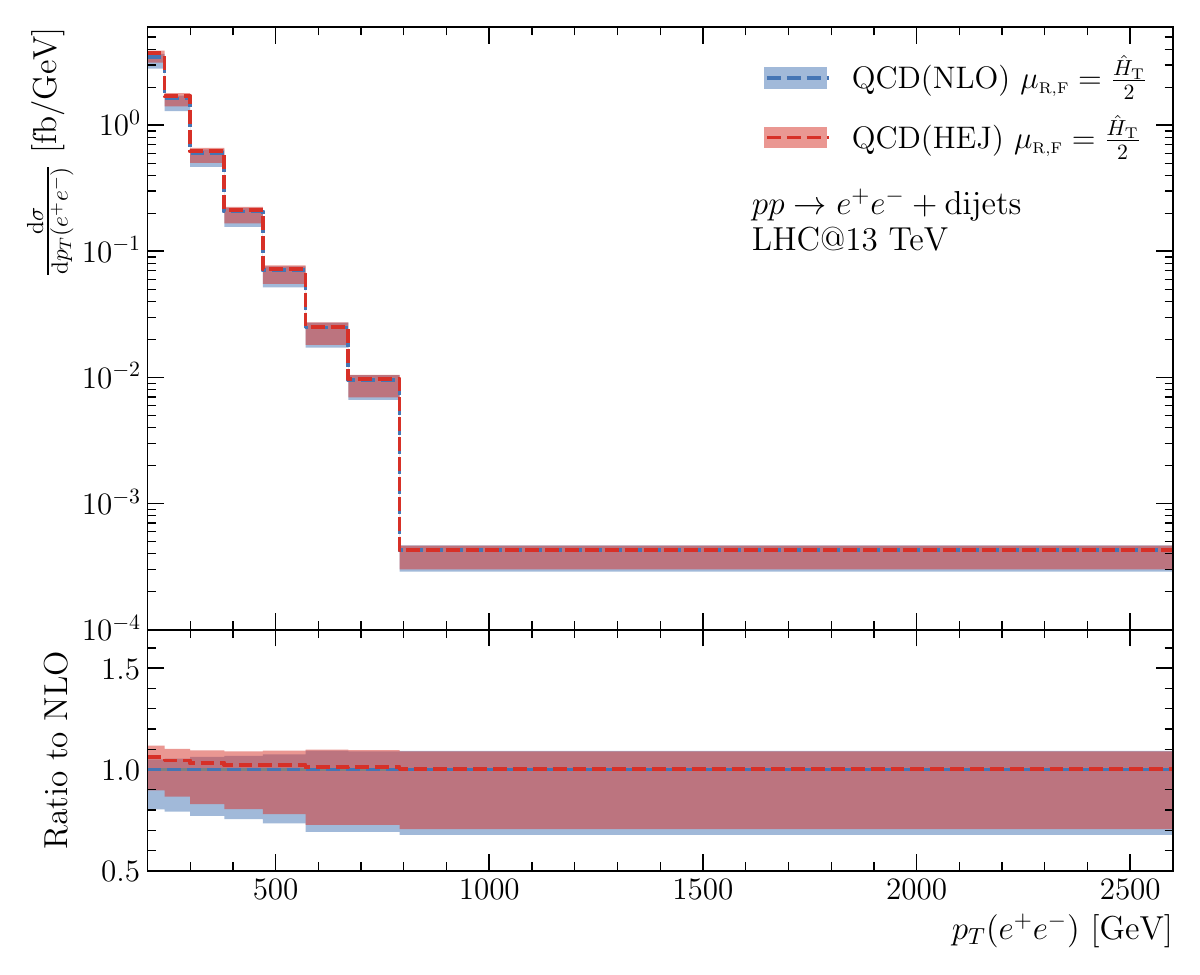}
  \includegraphics[width=0.48\textwidth]{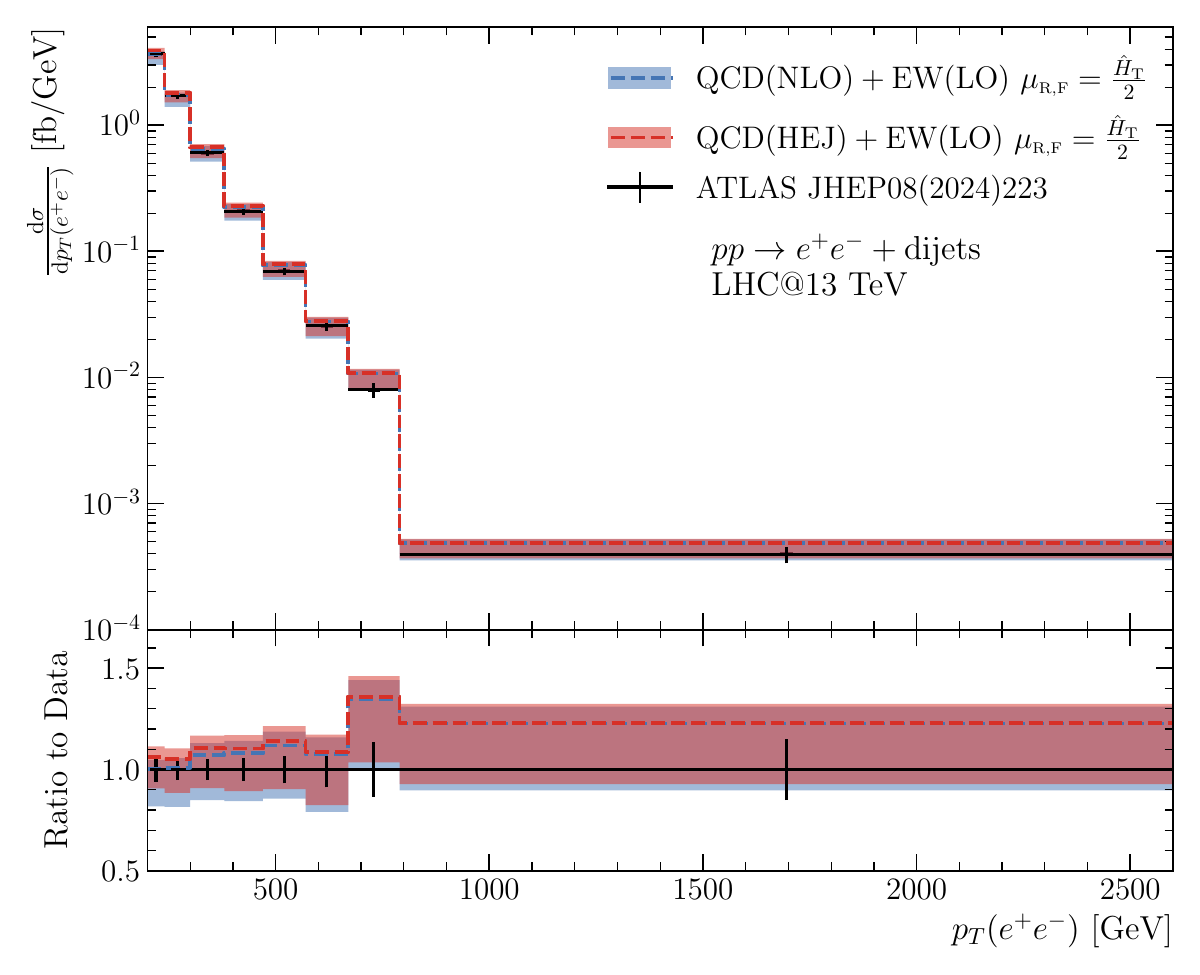}
  \caption{Left: $\di\sigma/\di p_{T}(e^+e^-)$ for \HEJ and NLO QCD for $pp\to jj
    (Z\to)\epem$(Born: $\as^2 \alpha^2$) with central scale choice
    $\mu_R=\mu_F=\htt$ and a 7-point scale variation band. Right: the same predictions with leading-order electroweak  contributions added and data points taken from~\cite{ATLAS:2024vqf}.}
  \label{fig:NLOHEJptV}
\end{figure}

Figure~\ref{fig:NLOHEJptV} compares the predictions for the $\ds/\di p_T(\epem)$ distribution.
The left panel shows the \HEJ (red) and NLO (blue) predictions for the QCD component.
The prediction from \HEJ here is very similar to that for NLO reflecting the fact that the matching corrections are modest.
This is a direct result of the fact that for these observables in general, \mjj and the high-energy effects are small.
The right panel adds the leading-order electroweak component to both predictions and compares to the data points in~\cite{ATLAS:2024vqf}.
Both the NLO (blue) and \HEJ (red) predictions, obtained with the central scale choice of $\mur=\muf=\Hhatt/2$, provide a good description of the distribution.

However, a closer inspection of the scale variation of the NLO predictions reveals this to be very asymmetrical.
This is then inherited by the \HEJ predictions through the matching procedure
(the \HEJ predictions do not show this asymmetry before matching, which can also be seen in figure~\ref{fig:HEJ} with similar variations up and down relative to the central result).
The NLO central line is so close to the upper band of the scale variation that it would seem implausible that the scale variation should provide any reasonable estimate of missing higher order corrections.
In fact the highly asymmetrical scale variation turns out to be a direct consequence of the increasingly large scale variation in the NLO prediction as a function of \mjj, as seen in figure~\ref{fig:fo}.
Indeed we checked that if~\mjj is constrained to the region below 1TeV then a more balanced scale variation at NLO is obtained for both $\ds/\di p_T(\epem)$ and $\ds/\di\Delta\phi_{jj}$.
In this sense the fixed order problem caused by the high-energy logarithms is masquerading and causing highly asymmetrical scale variations even in distributions where a central scale choice gives a good description of data.



\begin{figure}[tb]
  \centering
  \includegraphics[width=0.48\textwidth]{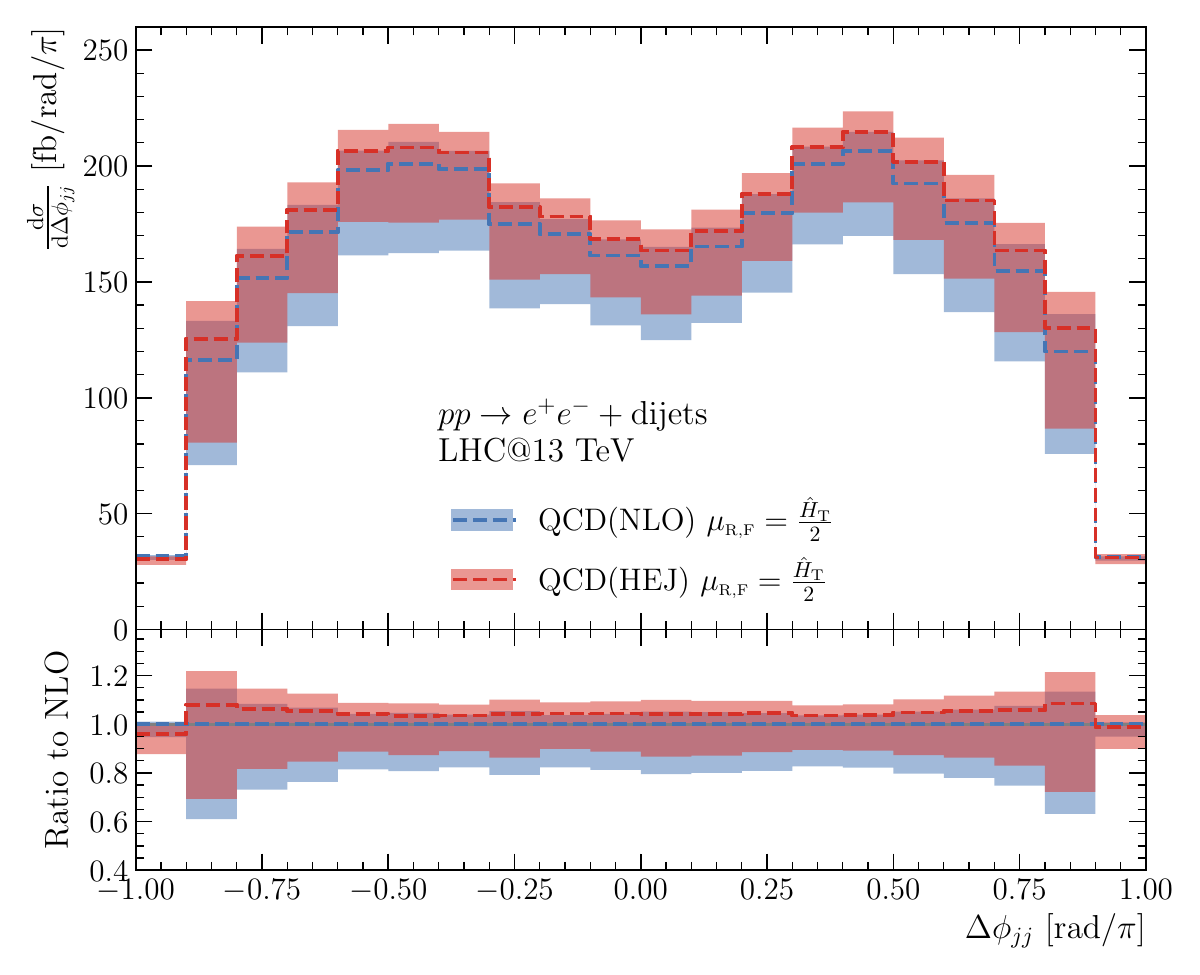}
  \includegraphics[width=0.48\textwidth]{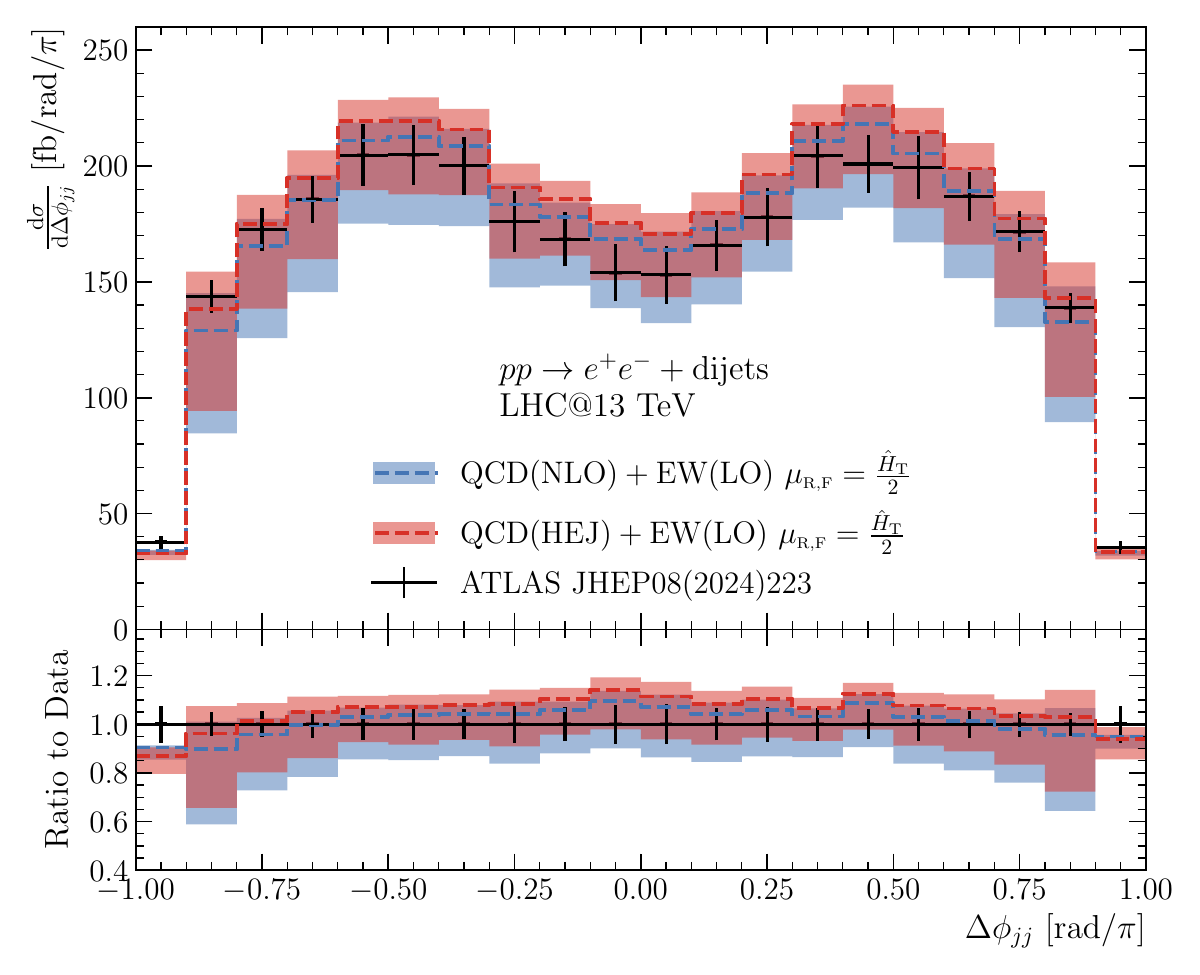}
  \caption{Left: $\di\sigma/\di \Delta\phi_{jj}$ for \HEJ and for NLO QCD for
    $pp\to jj (Z\to)\epem$(Born $\as^2 \alpha^2$) with central scale choice
    $\mu_R=\mu_F=\htt$ and a 7-point scale variation band. Right: the same predictions with leading-order electroweak  contributions added and data points taken from~\cite{ATLAS:2024vqf}.}
  \label{fig:NLOHEJphijj}
\end{figure}
Figure~\ref{fig:NLOHEJphijj} compares the two perturbative predictions for the $\ds/\di\Delta \phi_{jj}$ distribution, again with pure QCD predictions in the left panel and with leading-order electroweak corrections and data points added on the right.
Similar conclusions to those presented for $\ds/\di p_{T}(e^+e^-)$ apply:
the central prediction for NLO is good but at the upper edge of the results obtained by a standard 7-point scale variation.
The central prediction obtained by \HEJ is similar to that obtained at NLO, with the asymmetrical scale variation band inherited from NLO through the matching procedure.

The fact that the predictions obtained with \HEJ matched to NLO are so similar to the pure \NLO when the high-energy influence is small is an accolade to the matching procedure developed in~\cite{Andersen:2020yax}.
The issue of asymmetric scale variation is caused by the highly asymmetric scale variation in the \NLO results both at full \NLO and the \NLO high-energy approximation.
Ultimately, the impact of the fixed-order termination of the all-order description will be lessened by including high-order corrections, e.g.~NNLO.

\HighEJ and the setup used for producing all the results discussed in this paper are available at \href{http://hej.hepforge.org}{http://hej.hepforge.org}.

\section{Conclusions}
\label{sec:Conclusions}
We have exposed the impact of the high-energy logarithm on fixed order predictions.
The high-energy logarithms manifest themselves in several ways: the breakdown of the fixed order perturbative series, asymmetric scale variance and the need to choose an unusually large renormalisation scale to achieve numerically meaningful results at the cost of maximising the argument of a UV logarithm to counterbalance the genuine high-energy logarithm.
We demonstrated that the choice of renormalisation scale \mjj profoundly modifies the perturbative series at high energies and even removes the dependence on \as itself.

We systematically include the leading high-energy logarithmic corrections and demonstrate that this cures the problems seen at fixed order.
Specifically, the description is applied to the process $pp\to JJZ$.
The predictions presented here update the precision of the description of $pp\to ZJJ$ (in terms of both the logarithmic components and the fixed-order matching) to that used for the recent description of $pp\to \gamma JJ$~\cite{Andersen:2025zxw}.
We compare predictions for this process to data obtained by ATLAS~\cite{ATLAS:2024vqf} and demonstrate that the  \HighEJ framework describes correctly the slope in $\ds/\di\mjj$ and obtains a good description of distributions in both \mjj and the other two observables investigated, namely $\ds/\di p_T(\epem)$ and $\ds/\di\Delta\phi_{jj}$.

The predictions obtained with \HEJ achieve a good description of data using the standard renormalisation scale $\mur=\htt$ throughout.
This is contrary to the choices needed in fixed order where the choice of \htt is preferred for $\ds/\di p_T(\epem)$ and $\ds/\di\Delta\phi_{jj}$ but leads to unphysical results for fixed order for $\ds/\di\mjj$.

The fixed order results at NLO and beyond can be augmented by the all-order treatment presented here to properly address the issues caused by the high-energy logarithm.

We expect the conclusions presented here to generalise to higher perturbative orders and that the all-order resummation will repair also the issues which will arise at NNLO.

\acknowledgments

We would like to thank G.~Salam for discussions on the scale choice made for fixed-order perturbative predictions at large dijet invariant masses, specifically the point of high energy logarithms potentially being the true physical origin of the large negative NLO coefficients which are currently widely absorbed into anomalously large scale choices.
We thank Nigel Glover for a discussion on the formalism of the  1-loop results discussed in section~\ref{sec:helogfromnlo}.
Finally we thank C.~G\"utschow for help in making Rivet manage all the data processing from the calculations involved in the predictions presented in this study.
The authors express their thanks to current and previous collaborators of \HEJ, in particular B.~Duclou\'e for implementing the \LLp amplitudes used in this study.
The investigations reported here were made possible by the efficient management of the computing resources provided by the IPPP, University of Durham, and the UK GridPP~\cite{GridPP:2006wnd,Britton:2009ser}.
The work of Jeppe R.~Andersen is supported by the STFC under grant ST/X003167/1.
S.~Jaskiewicz is supported by the Swiss National Science Foundation Ambizione grant PZ00P2\_223524.

\appendix

\boldmath
\section{The resummation of the high-energy logarithms}
\label{sec:corlarges}
\label{sec:comp-high-energy}
\unboldmath
The resummation of high-energy logarithms rests on the universal factorisation of the amplitude in the MRK limit.
This factorisation up to NLL accuracy has been shown explicitly for up to two-loop five-point and three-loop four-point amplitudes~\cite{Caola:2021izf,Falcioni:2021dgr,Buccioni:2024gzo,Abreu:2024xoh,DelDuca:2025vux}.
Schematically, for a given rapidity ordering and parton flavour configuration it can be written as
\begin{equation}
  \label{eq:M_fact}
  \mathcal{M} \xrightarrow{\text{MRK}} B \times V \times L,
\end{equation}
where $B$ corresponds to the high-energy approximation of the
process-dependent Born-level matrix element, whereas $V$ and $L$
consist of process-independent factors describing the real and virtual
corrections, respectively. The sum runs over all kinematic
configurations contributing at a given logarithmic accuracy. The LL
matrix elements for the production of a charged lepton pair with jets
were first derived in~\cite{Andersen:2016vkp}. Ordering particles by
ascending rapidity, the kinematic configurations contributing at LL
accuracy can be written as $f_a f_b \to f_a \dots f_b \times (\ell^+
\ell^-)$, where $f_a, f_b$ denote the incoming parton flavours and the
ellipses represent an arbitrary number of emitted gluons.
The notation $\times (\ell^+ \ell^-)$ indicates that the rapidities of the charged leptons are arbitrary.
Note that this emission is only possible if at least one of the partons $f_a, f_b$ is a quark or antiquark.
Furthermore, at LL accuracy the outgoing partons $f_a$ and $f_b$ are hard, such that there are always at least two jets.

In this work, we include for the first time kinematic configurations
arising at NLL accuracy. We denote the resulting accuracy as LL+, as
full NLL accuracy would additionally require including subleading
corrections to the matrix elements for configurations contributing
already at LL. We distinguish between three kinds of NLL
configurations. Firstly, a ``central'' quark-antiquark pair $(q\bar{q})$
can be produced between $f_a$ and $f_b$, that is
$f_a f_b \to f_a \dots (q \bar{q}) \dots f_b\times (\ell^+
\ell^-)$. As before, the parentheses indicate an arbitrary rapidity
ordering between the contained particles. Secondly, an ``extremal''
quark-antiquark pair can be created at either end of the rapidity
chain from an incoming gluon,
$g f_b \to (q \bar{q}) \dots f_b\times (\ell^+ \ell^-)$ and
$f_a g \to f_a \dots (q \bar{q})\times (\ell^+ \ell^-)$. In the third
set of NLL configurations, an ``unordered'' gluon $g$ may be emitted
outside the LL rapidity ordering, leading to
$f_a f_b \to g f_a \dots f_b\times (\ell^+ \ell^-)$ with a quark or
antiquark $f_a$, and
$f_a f_b \to f_a \dots f_b g\times (\ell^+ \ell^-)$ with a quark or
antiquark $f_b$. To avoid collinear divergences which would only
cancel upon inclusion of the corresponding NLL virtual corrections,
we require the quark and the antiquark to be hard and form separate
jets. The same holds for unordered gluons. Hence, the central
quark-antiquark configurations first contribute to four-jet production
and the remaining NLL configurations to inclusive three-jet
production.

\begin{figure}[tb]
  \centering
  \begin{subfigure}{0.475\textwidth}
    \centering
    \includegraphics{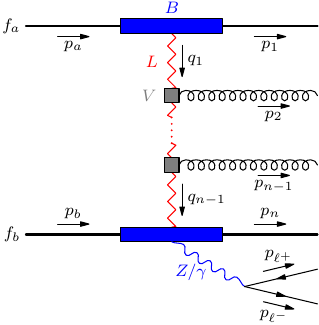}
    \caption{Leading logarithmic}
    \label{fig:HEJ_LL}
  \end{subfigure}
  \begin{subfigure}{0.475\textwidth}
  \centering
    \includegraphics{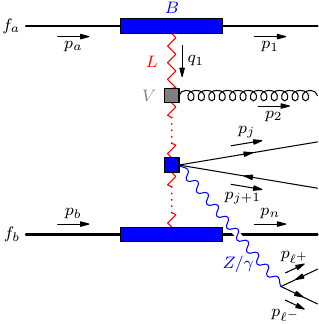}
    \caption{Central quark-antiquark}
    \label{fig:HEJ_cenqq}
  \end{subfigure}
  \begin{subfigure}{0.475\textwidth}
  \centering
    \includegraphics{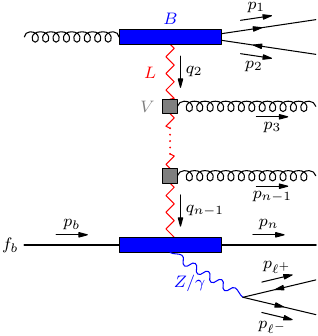}
    \caption{Backward quark-antiquark}
    \label{fig:HEJ_backwardqq}
  \end{subfigure}
  \begin{subfigure}{0.475\textwidth}
    \centering
    \includegraphics{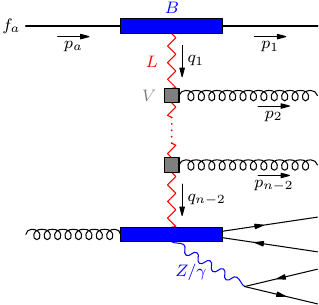}
    \caption{Forward quark-antiquark}
    \label{fig:HEJ_forwardqq}
  \end{subfigure}
  \begin{subfigure}{0.475\textwidth}
    \centering
    \includegraphics{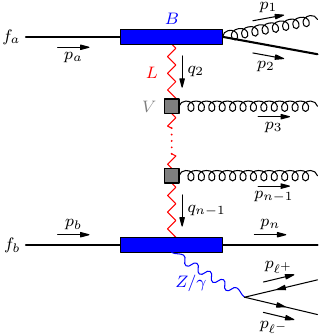}
    \caption{Backward unordered gluon}
    \label{fig:HEJ_unobackward}
  \end{subfigure}
  \begin{subfigure}{0.475\textwidth}
    \centering
    \includegraphics{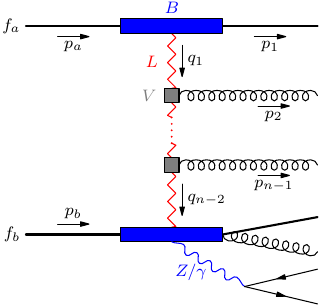}
    \caption{Forward unordered gluon}
    \label{fig:HEJ_unoforward}
  \end{subfigure}
  \caption{Matrix element structures for the included kinematic contributions. Outgoing partons are ordered by rapidity, while the lepton rapidities are arbitrary. Not shown, but included, are analogous configurations with flipped quark-antiquark ordering and contributions with charged lepton emission off other quark lines.}
  \label{fig:HEJ_ME}
\end{figure}

The LL and NLL configurations and corresponding matrix element structures are illustrated in figure~\ref{fig:HEJ_ME}.
Note that in the configurations involving an additional quark-antiquark pair the
matrix elements have to account for the possibility of this pair emitting the charged lepton pair.
Explicit expressions for the \HEJ matrix elements are obtained from the ones for the production of a W~boson with jets derived in~\cite{Andersen:2020yax}.
The only change required is to replace the W~propagator and couplings by the sum of the photon and Z~boson ones.
In contrast to the W~case, special care must be taken to account for the interference between contributions with the vector boson coupling to different quark lines when squaring the amplitude.
Hence, the modulus square of the \HEJ matrix element for a given kinematic configuration is given by
\begin{equation}
  \label{eq:M_HEJ}
  \lvert \mathcal{M}_{\text{\HEJ}} \rvert^2 = \sum_{e, e'} \mathcal{B}^{(e) (e')} \mathcal{V}^{(e) (e')} \mathcal{L}^{(e) (e')},
\end{equation}
where $\mathcal{B}^{(e) (e')}$ comprises the Born-level prediction, $\mathcal{V}^{(e)(e')}$ the all-order real-emission corrections, and $\mathcal{L}^{(e)(e')}$ the virtual corrections.
The sum runs over all lepton-pair emission sites that are allowed for the respective configuration.

To arrive at succinct formulas for $\mathcal{B}^{(e)(e')}$, $\mathcal{V}^{(e)(e')}$, and $\mathcal{L}^{(e)(e')}$, we denote the backward incoming parton momentum by $p_a$, the forward incoming parton momentum by $p_b$, and the $n$ final-state parton momenta by $p_1,\dots,p_n$, with corresponding rapidities $y_1 < \dots < y_n$.
The lepton momenta are $p_{\ell^-}$ and $p_{\ell^+}$ and we use the shorthand notation $p_{\gamma/Z} = p_{\ell^-} + p_{\ell^+}$ for the virtual vector boson momentum.
We label the possible vector boson coupling sites $e,e'$ by the number of the emitting parton, e.g.\ $e,e' \in \{1, n\}$ for LL configurations with two outgoing quarks or antiquarks.
For a vector boson coupling to a quark-antiquark pair at positions $j$ and $j+1$ we arbitrarily choose $j$ as the emission site label.
The final results for $\mathcal{B}^{(e)(e')}$, $\mathcal{V}^{(e)(e')}$, and $\mathcal{L}^{(e)(e')}$ do not depend on this choice.
We further introduce $q_i^{(e)}$ as the $i$th $t$-channel momentum, i.e.
\begin{equation}
  \label{eq:q}
  q_i^{(e)} = p_a - \sum_{j=1}^i p_j - \theta(i-e) (p_{\ell^-} + p_{\ell^+}),
\end{equation}
where $\theta$ is the Heaviside step function, and use the shorthand
notation $t^{(e)}_i = \left(q_i^{(e)}\right)^2$.

With these definitions, the all-order regulated virtual corrections read
\begin{align}
  \label{eq:virt}
  \mathcal{L}^{(e)(e')} ={}& \prod_{\substack{i=i_0\\i\ne j}}^{i_n - 1} \exp\left[\omega_0\left(q_{i\perp}^{(e)} q_{i\perp}^{(e')}\right) (y_{i+1} - y_i)\right],\displaybreak[0]\\
  \omega_0(q_\perp^2) ={}& - C_A \frac{\alpha_s}{\pi} \log \frac{q_\perp^2}{\lambda^2}
\end{align}
with $i_0 = 2$ for backward quark-antiquark or unordered gluon configurations and $i_0 = 1$ for all other configurations.
Similarly, $i_n = n - 1$ for forward quark-antiquark or unordered gluon configurations and otherwise $i_n = n$.
For a configuration with a central quark-antiquark pair at sites $j$, $j+1$ the corresponding factor is omitted.
The parameter $\lambda$ in the expression for the regularised Regge trajectory $\omega_0$ is an infrared regulator, see~\cite{Andersen:2017kfc} for details.
$C_A = 3$ is the usual quadratic Casimir invariant.

The regulated real-emission correction is given by
\begin{equation}
  \label{eq:real}
  \mathcal{V}^{(e)(e')} = \prod_{\substack{i=i_0+1\\i\notin \{j, j+1\}}}^{i_n-1} C_A g_s^2 \left(- \frac{V_\mu\left(q_{i-1}^{(e)},q_i^{(e)}\right) V^\mu\left(q_{i-1}^{(e')},q_i^{(e')}\right)}{\sqrt{t_{i-1}^{(e)}t_i^{(e)}t_{i-1}^{(e')}t_i^{(e')}}} - \theta(\lambda-p_{i\perp})\frac{4}{p_{i\perp}^2}\right).
\end{equation}

For the Lipatov vertex $V$ we use the expression
\begin{equation}
  \label{eq:V_Lipatov}
  \begin{split}
    V^\mu(q_{i-1}, q_i)={}& -(q_{i-1}+q_i)^\mu \\
    &+ \frac{p_a^\mu}{2} \left( \frac{q_{i-1}^2}{p_i\cdot p_a} +
      \frac{p_i\cdot p_b}{p_a\cdot p_b} + \frac{p_i\cdot p_n}{p_a\cdot p_n}\right) +
      (p_a \leftrightarrow p_1) \\
    &- \frac{p_b^\mu}{2} \left( \frac{q_i^2}{p_i \cdot p_b} + \frac{p_i\cdot
      p_a}{p_b\cdot p_a} + \frac{p_i\cdot p_1}{p_b\cdot p_1} \right)
      - (p_b \leftrightarrow p_n),
  \end{split}
\end{equation}
except for the unordered gluon configurations. For those, we instead use equation~(\ref{eq:V_Lipatov}) with the replacement $p_1 \to p_2$ for the backward case and $p_n \to p_{n-1}$ for the forward case.

Finally, the Born-level function $\mathcal{B}$ depends on the configuration.
The general structure reads
\begin{equation}
  \label{eq:B}
  \mathcal{B}^{(e)(e')} = g_s^{2n_B}\frac{K_{f_a}(p_a, p_1) K_{f_b}(p_b, p_n)}{4(N_c^2-1)} \frac{\mathcal{J}^{(e)(e')}}{\sqrt{t_{i_0}^{(e)} t_{i_n-1}^{(e)}t_{i_0}^{(e')}t_{i_n-1}^{(e')}}},
\end{equation}
where $n_B$ is the number of outgoing partons at Born level, namely $n_B=2$ for leading-logarithmic configurations, $n_B=3$ for backward or forward unordered gluon and quark-antiquark configurations, and $n_B=4$ for central quark-antiquark configurations.
We have further introduced
\begin{align}
  \label{eq:CAM_q}
  K_q(p, p') ={}& K_{\bar{q}}(p, p') = C_F,\displaybreak[0]\\
  \label{eq:CAM_g}
  K_g(p_a, p_1) ={}& \frac{1}{2}\left(\frac{p_1^-}{p_a^-} + \frac{p_a^-}{p_1^-}\right)\left(C_A - \frac{1}{C_A}\right) + \frac{1}{C_A},\displaybreak[0]\\
  K_g(p_b, p_n) ={}& \frac{1}{2}\left(\frac{p_n^+}{p_b^+} + \frac{p_b^+}{p_n^+}\right)\left(C_A - \frac{1}{C_A}\right) + \frac{1}{C_A},\displaybreak[0] \\
  \label{eq:CAM_gsplit}
  K_{g\to q\bar{q}}(p, p') ={}& \frac{1}{2}.
\end{align}
Here $K_{g\to q\bar{q}}$ applies to an incoming gluon which splits into the extremal quark-antiquark pair in place of $K_g$, c.f.\ figure~\ref{fig:HEJ_backwardqq} and \ref{fig:HEJ_forwardqq}.
$\mathcal{J}$ is a helicity sum over contracted currents. For LL configurations at most one of the incoming partons may be a gluon.
The allowed vector boson coupling sites are then $e = e' = 1$ if $f_b$ is a gluon, $e = e' = n$ if $f_a$ is a gluon, and $e,e' \in \{1,n\}$ if both are quarks or antiquarks.
One obtains the current contraction
\begin{align}
  \label{eq:j_contr_LL}
  \mathcal{J}^{(e)(e')}_{f_a f_b \to f_a f_b \times (\ell^+ \ell^-)} ={}& \sum_{\lambda_a,\lambda_b,\lambda_\ell}[j_{\ell^+\ell^-}\cdot j]^{(e)\lambda_a \lambda_b \lambda_\ell} \left([j_{\ell^+\ell^-}\cdot j]^{(e')\lambda_a \lambda_b \lambda_\ell}\right)^*,\displaybreak[0]\\
  \label{eq:j_contr_LL_Zbackward}
  [j_{\ell^+\ell^-}\cdot j]^{(1)\lambda_a \lambda_b \lambda_\ell} ={}& j_{\ell^+\ell^-}^{\lambda_a \lambda_\ell}(p_a, p_1) \cdot j^{\lambda_b}(p_b, p_n), \displaybreak[0]\\
  \label{eq:j_contr_LL_Zforward}
  [j_{\ell^+\ell^-}\cdot j]^{(n)\lambda_a \lambda_b \lambda_\ell} ={}& j_{\ell^+\ell^-}^{\lambda_b \lambda_\ell}(p_b, p_n) \cdot j^{\lambda_a}(p_a, p_1).
\end{align}
The (effective) currents are given by~\cite{Andersen:2012gk,Andersen:2016vkp,Andersen:2020yax}
\begin{align}
  \label{eq:j}
  j^{\mu \lambda}(p_a, p_1) ={}& \bar{u}^\lambda(p_1)\gamma^\mu u^\lambda(p_a),\displaybreak[0]\\
  \label{eq:jZ}
  j_{\ell^+\ell^-}^{\mu \lambda_q \lambda_\ell}(p_a, p_1) ={}& \mathcal{D}_{Z\gamma} j_\nu^{\lambda_\ell}(p_{\ell^+}, p_{\ell^-})\bar{u}^{\lambda_q}(p_1)\left(
  \gamma^\nu \frac{\slashed{p}_1 + \slashed{p}_{\gamma/Z}}{(p_1 + p_{\gamma/Z})^2} \gamma^\mu
    + \gamma^\mu \frac{\slashed{p}_a - \slashed{p}_{\gamma/Z}}{(p_a - p_{\gamma/Z})^2} \gamma^\nu\right) u^{\lambda_q}(p_a),
        \displaybreak[0]\\
    \mathcal{D}_{Z\gamma} ={}& -g^2 \left(\frac{1}{c_w^2}\frac{(T_{3q}^{\lambda_q} - Q_q s_w^2)(T_{3\ell}^{\lambda_\ell} - Q_\ell s_w^2)}{p_{\gamma/Z}^2 - M_Z^2 + i\Gamma_Z M_Z} + Q_q Q_\ell \frac{s_w^2}{p_{\gamma/Z}^2}\right).
\end{align}
We have denoted the weak coupling constant by $g$ and the sine and cosine of the weak mixing angle by $s_w$ and $c_w$, respectively.
$T_{3f}^{\lambda_f}$ and $Q_f$ are the weak isospin and electric charge of a fermion $f$ with helicity $\lambda_f$. $M_Z$ and $\Gamma_Z$ are the mass and width of the Z boson.

\subsection{Current contractions for inclusive three-jet NLL configurations}
\label{sec:NLL_uno_exqqbar}

For the NLL configurations the dependence of the current contraction on the vector boson coupling site is more involved.
For a forward unordered gluon configuration with a gluon momentum $p_g \equiv p_n$ we have~\cite{Andersen:2020yax}
\begin{align}
  \label{eq:j_contr_unob}
  \mathcal{J}^{(e)(e')}_{f_a f_b \to f_a f_b g \times (\ell^+ \ell^-)} ={}& \sum_{\substack{\lambda_a,\lambda_b,\\ \lambda_\ell,\lambda_g}} \left[ C_F \left(X^{(e)*} X^{(e')} + Y^{(e)*} Y^{(e')}\right) - \frac{1}{2C_A} \left(X^{(e)*} Y^{(e')} + Y^{(e)*} X^{(e')}\right)\right],\displaybreak[0]\\
  \label{eq:Xdef}
  X^{(e)} ={}& U^{(e)}_1 - L^{(e)}, \displaybreak[0]\\
  \label{eq:Ydef}
  Y^{(e)} ={}& U^{(e)}_2 + L^{(e)},
\end{align}
where we have left the dependence of $X, Y, U, L$ on the four helicities $\lambda_a,\lambda_b,\lambda_\ell,\lambda_g$ implicit for the sake of readability.
If the backward incoming parton $f_a$ is a quark or antiquark, the vector boson can couple to this fermion line.
For this case, we obtain
\begin{align}
  \left(U^{(1)}_1, U^{(1)}_2, L^{(1)}\right) = {}& j_{\ell^+\ell^-}^{\mu\lambda_a \lambda_\ell}(p_a, p_1) \epsilon^{\nu \lambda_g}(p_g) \left(U^{(1)}_{1\mu\nu}, U^{(1)}_{2\mu\nu}, L^{(1)}_{\mu\nu}\right),\displaybreak[0]\\
  U^{(1)}_{1\mu\nu} ={}& \frac{1}{(p_{n-1} + p_g)^2} \left[2p_{{n-1}\nu}  j^{\lambda_b}_\mu(p_b, p_{n-1}) + j^{\lambda_b}_\nu(p_g, p_{n-1})j^{\lambda_b}_\mu(p_b, p_g)\right],\displaybreak[0]\\
  U^{(1)}_{2\mu\nu} ={}& \frac{1}{(p_b - p_g)^2} \left[2p_{b\nu}  j^{\lambda_b}_\mu(p_b, p_{n-1}) - j^{\lambda_b}_\mu(p_g, p_{n-1})j^{\lambda_b}_\nu(p_b, p_g)\right],\displaybreak[0]\\
  L^{(1)}_{\mu\nu} ={}& \frac{1}{(p_{n-1}-p_b)^2}\bigg[2 p_g \cdot j^{\lambda_b}(p_b, p_{n-1}) g_{\mu\nu} -2p_{g\mu}  j^{\lambda_b}_\nu(p_b, p_{n-1}) \nonumber\\
  & + t_{n-2}^{(1)} j^{\lambda_b}_\mu(p_b, p_{n-1}) \left(\frac{p_{1\nu}}{(p_1 + p_g)^2}
  + \frac{p_{a\nu}}{(p_a + p_g)^2}\right)  \nonumber\\
&+\left(p_g - 2q_{n-2}^{(1)}\right)_\nu j^{\lambda_b}_\mu(p_b, p_{n-1})\bigg].
\end{align}
For forward unordered configurations the forward incoming parton $f_b$ is always a quark or antiquark. The contribution where the vector boson couples to this line yields
\begin{align}
  \left(U^{(n-1)}_1, U^{(n-1)}_2, L^{(n-1)}\right) = {}& \mathcal{D}_{Z\gamma} j^{\mu\lambda_1}(p_a, p_1) \epsilon^{\nu \lambda_g}(p_g) j^{\rho\lambda_\ell}(p_{\ell^+},p_{\ell^-})\left(U^{(n-1)}_{1\mu\nu\rho}, U^{(n-1)}_{2\mu\nu\rho}, L^{(n-1)}_{\mu\nu\rho}\right),\displaybreak[0]\\
  U^{(n-1)}_{1\mu\nu\rho} ={}& \phantom{{}+{}}\bar{u}^{\lambda_b}(p_{n-1})\gamma_\nu\frac{\slashed{p}_{n-1}+\slashed{p}_g}{(p_{n-1}+p_g)^2}\gamma_\mu \frac{\slashed{p}_b - \slashed{p}_{\gamma/Z}}{(p_b - p_{\gamma/Z})^2} \gamma_\rho u^{\lambda_b}(p_b)\nonumber\\
  &+\bar{u}^{\lambda_b}(p_{n-1})\gamma_\nu\frac{\slashed{p}_{n-1}+\slashed{p}_g}{(p_{n-1}+p_g)^2}\gamma_\rho \frac{\slashed{p}_{n-1}+\slashed{p}_g + \slashed{p}_{\gamma/Z}}{(p_{n-1}+p_g + p_{\gamma/Z})^2} \gamma_\mu u^{\lambda_b}(p_b)\nonumber\\
  &+\bar{u}^{\lambda_b}(p_{n-1})\gamma_\rho\frac{\slashed{p}_{n-1}+\slashed{p}_{\gamma/Z}}{(p_{n-1}+p_{\gamma/Z})^2}\gamma_\nu \frac{\slashed{p}_{n-1}+\slashed{p}_g + \slashed{p}_{\gamma/Z}}{(p_{n-1}+p_g + p_{\gamma/Z})^2} \gamma_\mu u^{\lambda_b}(p_b),\displaybreak[0]\\
  U^{(n-1)}_{2\mu\nu\rho} ={}& \phantom{{}+{}}\bar{u}^{\lambda_b}(p_{n-1})\gamma_\mu\frac{\slashed{p}_b-\slashed{p}_{\gamma/Z} - \slashed{p}_g}{(p_b-p_{\gamma/Z} - p_g)^2}\gamma_\nu \frac{\slashed{p}_b - \slashed{p}_{\gamma/Z}}{(p_b - p_{\gamma/Z})^2} \gamma_\rho u^{\lambda_b}(p_b)\nonumber\\
  &+\bar{u}^{\lambda_b}(p_{n-1})\gamma_\mu\frac{\slashed{p}_b-\slashed{p}_{\gamma/Z} - \slashed{p}_g}{(p_b-p_{\gamma/Z} - p_g)^2}\gamma_\rho \frac{\slashed{p}_b - \slashed{p}_g}{(p_b - p_g)^2} \gamma_\nu u^{\lambda_b}(p_b)\nonumber\\
  &+\bar{u}^{\lambda_b}(p_{n-1})\gamma_\rho\frac{\slashed{p}_{n-1}+\slashed{p}_{\gamma/Z}}{(p_{n-1}+p_{\gamma/Z})^2}\gamma_\mu \frac{\slashed{p}_b - \slashed{p}_g}{(p_b - p_g)^2} \gamma_\nu u^{\lambda_b}(p_b),\displaybreak[0]\\
L^{(n-1)}_{\mu\nu\rho}={}&\bar{u}^{\lambda_b}(p_{n-1})\left(\gamma_\sigma\frac{\slashed{p}_b-\slashed{p}_{\gamma/Z}}{(p_b-p_{\gamma/Z})^2}\gamma_\rho  + \gamma_\rho\frac{\slashed{p}_{n-1}+\slashed{p}_{\gamma/Z}}{(p_{n-1}+p_{\gamma/Z})^2}\gamma_\sigma\right) u^{\lambda_b}(p_b)\nonumber\\
&\times \bigg\{(p_{g} + q^{(n-1)}_{n-2})_\sigma  g_{\mu\nu} - 2p_{g\mu}g_{\nu\sigma}\nonumber\\
&\ +\left[\left(\frac{p_a}{(p_a + p_g)^2} + \frac{p_1}{(p_1 + p_g)^2}\right) t^{(n-1)}_{n-2} - 2 q^{(n-1)}_{n-2} + p_{g}\right]_\nu g_{\mu\sigma}\bigg\}\nonumber\\
&\times \frac{1}{(p_b - p_{n-1} - p_{\gamma/Z})^2}.
\end{align}
The current contraction for a backward unordered gluon configuration is completely analogous.
Configurations with backward or forward quark-antiquark pairs are obtained from the corresponding unordered gluon configuration via crossing symmetry, i.e.\ by replacing $p_b \leftrightarrow -p_n$ for the forward and $p_a \leftrightarrow -p_1$ for the backward case~\cite{Andersen:2020yax}.
The modulus square of the matrix element does not depend on the relative rapidity ordering between the quark and the antiquark.

\subsection{Current contractions for central quark-antiquark configurations}
\label{sec:NLL_cenqqbar}

In the remaining configurations with a central quark-antiquark pair at rapidity positions $j, j+1$ the current contraction involves two currents for the most forward and backward partons as well as an effective quark-antiquark emission vertex.
The general structure is the same as for the other NLL configurations, c.f.\ equation~\eqref{eq:j_contr_unob}:
\begin{equation}
  \label{eq:j_contr_cenqqbar}
  \mathcal{J}^{(e)(e')}_{f_a f_b \to f_a (q\bar{q}) f_b \times (\ell^+ \ell^-)} = \frac{C_F}{16} \sum_{\substack{\lambda_a,\lambda_b,\\ \lambda_\ell,\lambda_q}} \frac{2 C_F C_A \left(V^{(e)*} V^{(e')} + W^{(e)*} W^{(e')}\right) + V^{(e)*} W^{(e')} + W^{(e)*} V^{(e')}}{\sqrt{t_{j-1}^{(e)} t_{j+1}^{(e)} t_{j-1}^{(e')} t_{j+1}^{(e')}}}.
\end{equation}
Here, $V$ and $W$ are contractions of two currents with an effective
vertex for quark-antiquark emission. The vector coupling sites $e,e'$
can take the value $j$ for coupling to the central quark-antiquark
pair, $1$ if $f_a$ is a quark or antiquark, and $n$ if $f_b$ is a
quark or antiquark. Hence, we find
\begin{align}
  V^{(1)} ={}& j^{\lambda_a \lambda_\ell}_{\ell^+\ell^-}(p_a, p_1) \cdot \left(X_s^{\lambda_q} + X_6^{\lambda_q}\right) \cdot j^{\lambda_b}(p_b,p_n),
  \\
    V^{(j)} ={}& j^{\lambda_a}(p_a, p_1) \cdot  \mathcal{D}_{Z\gamma}\left(X_{s \ell^+\ell^-}^{\lambda_q\lambda_{\ell}} + X_{6 \ell^+\ell^-}^{\lambda_q\lambda_{\ell}}\right) \cdot j^{\lambda_b}(p_b,p_n),
  \\
  V^{(n)} ={}& j^{\lambda_a}(p_a, p_1) \cdot \left(X_s^{\lambda_q} + X_6^{\lambda_q}\right) \cdot j^{\lambda_b \lambda_\ell}_{\ell^+\ell^-}(p_b,p_n),\\
  W^{(1)} ={}& j^{\lambda_a \lambda_\ell}_{\ell^+\ell^-}(p_a, p_1) \cdot \left(X_s^{\lambda_q} + X_7^{\lambda_q}\right) \cdot j^{\lambda_b}(p_b,p_n),
  \\
    W^{(j)} ={}& j^{\lambda_a}(p_a, p_1) \cdot \mathcal{D}_{Z\gamma}\left(X_{s \ell^+\ell^-}^{\lambda_q\lambda_{\ell}} + X_{7 \ell^+\ell^-}^{\lambda_q\lambda_{\ell}}\right) \cdot j^{\lambda_b}(p_b,p_n),
  \\
  W^{(n)} ={}& j^{\lambda_a}(p_a, p_1) \cdot \left(X_s^{\lambda_q} + X_7^{\lambda_q}\right) \cdot j^{\lambda_b \lambda_\ell}_{\ell^+\ell^-}(p_b,p_n)
\end{align}
with the currents $j$ and $j_{\ell^+\ell^-}$ from eqs.~\eqref{eq:j}, \eqref{eq:jZ}. The vertices for quark-antiquark creation are given by~\cite{Andersen:2020yax}
\begin{align}
  \label{eq:Xs}
  X_s^{\mu\nu\lambda_q} ={}& \left(g^{\mu\nu} X_\text{sym}^\sigma + X_5^{\mu\nu\sigma}\right)\frac{j_\sigma^{\lambda_q}(p_{\bar{q}}, p_q)}{p_{q\bar{q}}^2},\displaybreak[0]\\
  \label{eq:Xsym}
  X_\text{sym}^\mu ={}& t_{j-1}^{(e)}\left(\frac{p_a^\mu}{2 p_a \cdot p_{q\bar{q}}} + \frac{p_1^\mu}{2 p_1 \cdot p_{q\bar{q}}}\right) - t_{j+1}^{(e)}\left(\frac{p_b^\mu}{2 p_b \cdot p_{q\bar{q}}} + \frac{p_n^\mu}{2 p_n \cdot p_{q\bar{q}}}\right),\displaybreak[0]\\
  \label{eq:X5}
  X_5^{\mu\nu\sigma} ={}& \left(q_{j-1}^{(e)} + p_{q\bar{q}}\right)^\nu g^{\mu\sigma}
                          + \left(q_{j+1}^{(e)} - p_{q\bar{q}}\right)^\mu g^{\nu\sigma}
                          - \left(q_{j-1}^{(e)} + q_{j+1}^{(e)}\right)^\sigma g^{\mu\nu},\displaybreak[0]\\
  \label{eq:X6}
  X_6^{\mu\nu\lambda_q}={}& \bar{u}^{\lambda_q}(p_q)\gamma^\mu\frac{\slashed{q}_{j-1}^{(e)} - \slashed{p}_q}{\left(q_{j-1}^{(e)} - p_q\right)^2}\gamma^\nu u^{\lambda_q}(p_{\bar{q}}),\displaybreak[0]\\
  \label{eq:X7}
  X_7^{\mu\nu\lambda_q}={}& \bar{u}^{\lambda_q}(p_q)\gamma^\nu\frac{\slashed{q}_{j-1}^{(e)} - \slashed{p}_{\bar{q}}}{\left(q_{j-1}^{(e)} - p_{\bar{q}}\right)^2}\gamma^\mu u^{\lambda_q}(p_{\bar{q}}).
\end{align}
To better exhibit the structure of the result we have defined $p_q \equiv p_j, p_{\bar{q}} \equiv p_{j+1}, p_{q\bar{q}} = p_q + p_{\bar{q}}$, noting that the choice $p_q \equiv p_{j+1}, p_{\bar{q}} \equiv p_j$ leads to the same result for the current contraction.
Finally, for the simultaneous production of a quark-antiquark and the lepton-antilepton pair we obtain~\cite{Andersen:2020yax}
\begin{align}
  \label{eq:Xsll}
 X_{s\ell^+\ell^-}^{\mu\nu\lambda_q\lambda_\ell} ={}& -i\left(g^{\mu\nu} X_{\text{sym}\ell^+\ell^-}^\sigma + X_{5\ell^+\ell^-}^{\mu\nu\sigma}\right)\frac{j_{\sigma\ell^+\ell^-}^{\lambda_q\lambda_\ell}(-p_{\bar{q}}, p_q)}{(p_{q\bar{q}} + p_{\gamma/Z})^2},\displaybreak[0]\\
  \label{eq:Xsymll}
  X_{\text{sym}\ell^+\ell^-}^\mu ={}& t_{j-1}^{(e)}\left(\frac{p_a^\mu}{2 p_a \cdot (p_{q\bar{q}} + p_{\gamma/Z})} + \frac{p_1^\mu}{2 p_1 \cdot (p_{q\bar{q}}+p_{\gamma/Z})}\right)\nonumber\\
  &- t_{j+1}^{(e)}\left(\frac{p_b^\mu}{2 p_b \cdot (p_{q\bar{q}} + p_{\gamma/Z})} + \frac{p_n^\mu}{2 p_n \cdot (p_{q\bar{q}} + p_{\gamma/Z})}\right),\displaybreak[0]\\
  \label{eq:X5ll}
  X_{5\ell^+\ell^-}^{\mu\nu\sigma} ={}& \left(q_{j-1}^{(e)} + p_{q\bar{q}} + p_{\gamma/Z}\right)^\nu g^{\mu\sigma}
+ \left(q_{j+1}^{(e)} - p_{q\bar{q}}  - p_{\gamma/Z}\right)^\mu g^{\nu\sigma}
- \left(q_{j-1}^{(e)} + q_{j+1}^{(e)}\right)^\sigma g^{\mu\nu},\displaybreak[0]\\
  \label{eq:X6ll}
  X_{6\ell^+\ell^-}^{\nu\mu\sigma\lambda_q\lambda_\ell}={}&-j_\sigma^{\lambda_\ell}(p_{\ell^+}, p_{\ell^-}) \bar{u}^{\lambda_q}(p_q)
\begin{aligned}[t]
  &\left[- \gamma^\sigma \frac{\slashed{p}_q + \slashed{p}_{\gamma/Z}}{(p_q + p_{\gamma/Z})^2}\gamma^\mu \frac{\slashed{q}^{(j)}_{j+1} + \slashed{p}_{\bar{q}}}{\left(q^{(j)}_{j+1} + p_{\bar{q}}\right)^2}\gamma^\nu \right.\\
        &\quad+ \gamma^\mu \frac{\slashed{q}_{j-1}^{(j)} - \slashed{p}_q}{\left(q^{(j)}_{j-1} - p_q\right)^2}\gamma^\sigma \frac{\slashed{q}^{(j)}_{j+1} + \slashed{p}_{\bar{q}}}{\left(q^{(j)}_{j+1} + p_{\bar{q}}\right)^2}\gamma^\nu\\
        &\quad\left. + \gamma^\mu \frac{\slashed{q}_{j-1}^{(j)} - \slashed{p}_q}{\left(q^{(j)}_{j-1} - p_q\right)^2}\gamma^\nu \frac{\slashed{p}_{\bar{q}}+\slashed{p}_{\gamma/Z}}{\left(p_{\bar{q}}+p_{\gamma/Z}\right)^2}\gamma^\sigma\right]u^{\lambda_q}(p_{\bar{q}}),
\end{aligned}
\displaybreak[0]\\
\label{eq:X7ll}
 X_{7\ell^+\ell^-}^{\nu\mu\sigma\lambda_q\lambda_\ell}={}& j_\sigma^{\lambda_\ell}(p_{\ell^+}, p_{\ell^-}) \bar{u}^{\lambda_q}(p_q)
\begin{aligned}[t]
  &\left[- \gamma^\nu \frac{\slashed{q}^{(j)}_{j+1} + \slashed{p}_q}{\left(q^{(j)}_{j+1} + p_q\right)^2}\gamma^\mu \frac{\slashed{p}_{\bar{q}}+\slashed{p}_{\gamma/Z}}{(p_{\bar{q}}+ p_{\gamma/Z})^2}\gamma^\sigma \right.\\
  &\quad+ \gamma^\nu \frac{\slashed{q}_{j+1}^{(j)} + \slashed{p}_q}{\left(q^{(j)}_{j+1} + p_q\right)^2}\gamma^\sigma \frac{\slashed{q}^{(j)}_{j-1} - \slashed{p}_{\bar{q}}}{\left(q^{(j)}_{j-1} - p_{\bar{q}}\right)^2}\gamma^\mu\\
  &\quad\left. + \gamma^\sigma \frac{\slashed{p}_q + \slashed{p}_{\gamma/Z}}{(p_q + p_{\gamma/Z})^2}\gamma^\nu \frac{\slashed{q}^{(j)}_{j-1} - \slashed{p}_{\bar{q}}}{\left(q^{(j)}_{j-1} - p_{\bar{q}}\right)^2}\gamma^\mu\right]u^{\lambda_q}(p_{\bar{q}}).
\end{aligned}
\end{align}
LL+ resummation is then added to leading-order samples by reweighting events with kinematics compatible with the configurations of figure~\ref{fig:HEJ_ME} by the corresponding all-order \HEJ matrix elements, see~\cite{Andersen:2017kfc,Andersen:2019yzo} for details.

Leading order samples of configurations which do not arise in the LL+ resummation are added up to the multiplicity of a given observable.
In this case up to two jets.
This ensures that the description of any observable at this stage is leading order accurate plus resummation.

\bibliographystyle{JHEP}
\bibliography{main.bib}

\end{document}